\documentclass[dvipsnames]{article}

\usepackage[final]{openbmb_2025}

\usepackage{amsmath}
\usepackage{amssymb}
\usepackage{mathtools}
\usepackage{amsthm}
\usepackage{multirow}
\usepackage{makecell}
\usepackage{subcaption}
\usepackage{wrapfig}
\usepackage{graphicx}
\usepackage{pifont}

\usepackage[utf8]{inputenc} % allow utf-8 input
\usepackage[T1]{fontenc}    % use 8-bit T1 fonts
\usepackage{url}            % simple URL typesetting
\usepackage{booktabs}       % professional-quality tables
\usepackage{amsfonts}       % blackboard math symbols
\usepackage{nicefrac}       % compact symbols for 1/2, etc.
\usepackage{microtype}      % microtypography

\usepackage{color, soul}
\usepackage[most,breakable]{tcolorbox}
\usepackage{xcolor}

\usepackage[labelfont=bf]{caption}
\usepackage{enumitem}
\usepackage{sectsty}

\usepackage{fancyhdr}
\renewcommand{\headrulewidth}{1pt}
\makeatletter
\def\headrule{{\if@fancyplain\let\headrulewidth\plainheadrulewidth\fi
\hrule\@height\headrulewidth\@width\textwidth \vskip-\headrulewidth}}
\makeatother

\definecolor{BMBDarkBlue}{HTML}{315EFE}
\definecolor{BMBLightBlue}{HTML}{00D3ED}
\usepackage[colorlinks, linkcolor=BMBDarkBlue, anchorcolor=BMBDarkBlue, citecolor=BMBDarkBlue, urlcolor=BMBDarkBlue]{hyperref}

\sectionfont{\color{BMBDarkBlue}\fontfamily{zi4}\selectfont}
\subsectionfont{\color{BMBDarkBlue}\fontfamily{zi4}\selectfont}
\subsubsectionfont{\color{BMBDarkBlue}\fontfamily{zi4}\selectfont}

\newtcolorbox{mytheorem}{
  colback=gray!5,       % BackGround Color
  colframe=gray!80,     % Frame Color
  boxrule=0.5pt,        % Frame Width
  arc=4pt,              % Fillet Radius
  left=4pt,             % Left Margin
  right=4pt,            % Right Margin
  top=4pt,              % Top Margin
  bottom=4pt,           % Bottom Margin
}

\newcommand{\fancyheadname}{\textit{\textbf{UltraEval-Audio}}}
\title{UltraEval-Audio: A Unified Framework for Comprehensive Evaluation of Audio Foundation Models}

\author{%
Qundong Shi\textsuperscript{\rm 1}\thanks{Equal Contributions.}, 
Jie Zhou\textsuperscript{\rm 1}\footnotemark[1],
Biyuan Lin\textsuperscript{\rm 1}, 
Junbo Cui\textsuperscript{\rm 1},
Guoyang Zeng\textsuperscript{\rm 1},
Yixuan Zhou\textsuperscript{\rm 1},\\
\textbf{
Ziyang Wang\textsuperscript{\rm 1},
Xin Liu\textsuperscript{\rm 1},
Zhen Luo\textsuperscript{\rm 1},
Yudong Wang\textsuperscript{\rm 2}\thanks{Corresponding Authors.}, 
Zhiyuan Liu\textsuperscript{\rm 2}\footnotemark[2]}\\
\textsuperscript{\rm 1}ModelBest Inc.~~ \textsuperscript{\rm 2}Tsinghua University\\
\texttt{\{shiqundong, zhoujie, linbiyuan\}@modelbest.cn}\\
\texttt{\{yudongwang, liuzy\}@tsinghua.edu.cn}
}

\def\github{\raisebox{-1.5pt}{\includegraphics[height=1.05em]{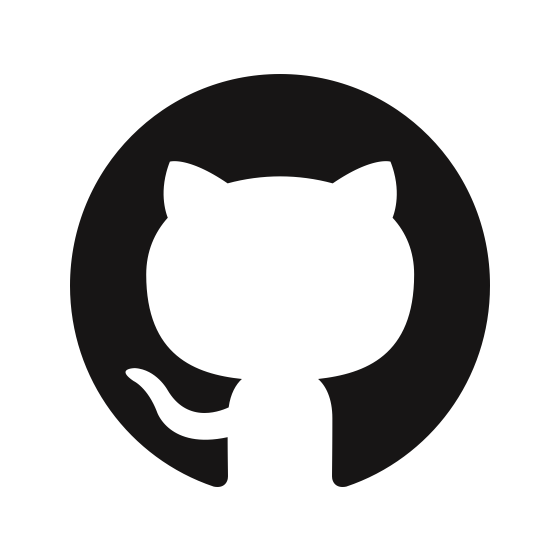}}}
\newcommand{\githublink}{https://github.com/OpenBMB/UltraEval-Audio}

\begin{document}

\maketitle
\thispagestyle{fancy} % 恢复封面页的页眉和页脚

\vspace{1em}

\begin{abstract}
%自 GPT-4o 问世以来，音频基础模型虽取得突破性进展，但其系统性评测，尤其是针对音频生成任务方面的缺失，已成为制约领域进一步发展的核心瓶颈。
The development of audio foundation models has accelerated rapidly since the emergence of GPT-4o. However, the lack of comprehensive evaluation has become a critical bottleneck for further progress in the field, particularly in audio generation. 
%当前音频评测体系面临三重严峻挑战：（1）作为模型底座的音频编解码器，其离散表征的多维性能表现长期缺乏系统性度量；（2）当前以用户为中心的语音评测基准以英文为主，导致无法客观评估模型在中文上的表现能力；（3）音频评测缺乏统一的框架，相关数据集和代码分散在不同来源，极大阻碍了模型间公平、高效的横向对比。
Current audio evaluation faces three major challenges: 
(1) audio evaluation lacks a unified framework, with datasets and code scattered across various sources, hindering fair and efficient cross-model comparison;
(2) audio codecs, as a key component of audio foundation models, lack a widely accepted and holistic evaluation methodology; 
(3) existing speech benchmarks are heavily reliant on English, making it challenging to objectively assess models' performance on Chinese.
%针对第一个问题，我们构建了一套针对音频编解码器的综合评估方法，从语义信息、音色保真度和声学质量三个方面进行全面分析； 针对第二个问题，我们提出了 SpeechCMMLU 和 SpeechHSK 两个全新的中文评测基准，用于评估中文知识能力和语言熟练度。 针对评测框架分散的问题，我们通过模块化的设计，构建了支持 10 种语言与 14 类任务的评测框架，整合了编解码器评估方案与中文基准，并系统性集成 24 个主流模型与 36 个权威基准。
To address the first issue, we introduce
% have developed 
\textbf{UltraEval-Audio}, a unified evaluation framework for audio foundation models, specifically designed for both audio understanding and generation tasks. UltraEval-Audio features a modular architecture, supporting 10 languages and 14 core task categories, while seamlessly integrating 24 mainstream models and 36 authoritative benchmarks. To enhance research efficiency, the framework provides a one-command evaluation feature, accompanied by real-time public leaderboards.
For the second challenge, UltraEval-Audio adopts a novel comprehensive evaluation scheme for audio codecs, evaluating performance across three key dimensions: semantic accuracy, timbre fidelity, and acoustic quality.
To address the third issue, we propose two new Chinese benchmarks, \textit{SpeechCMMLU} and \textit{SpeechHSK}, designed to assess Chinese knowledge proficiency and language fluency.
We wish that \textbf{UltraEval-Audio} will provide both academia and industry with a transparent, efficient, and fair platform for comparison of audio models.
Our code, benchmarks, and leaderboards are available at https://github.com/OpenBMB/UltraEval-Audio.
\end{abstract}

\vspace{-1.0em}
\section{Introduction}
\vspace{-1.0em}

Following the groundbreaking success of large language models (LLMs), the rapid development of multimodal large models has ensued. 
Notably, the release of OpenAI's GPT-4o~\citep{openai2024gpt4ocard} marked the advent of the native audio interaction era, accelerating the explosive development of audio foundation models (AFMs). Subsequently, a series of models with complex understanding and generation capabilities emerged, including Qwen2.5-Omni~\citep{xu2025qwen2}, Moshi~\citep{kyutai2024moshi}, GLM-4-Voice~\citep{zeng2024glm4voiceintelligenthumanlikeendtoend}, Step-Audio~\citep{huang2025stepaudiounifiedunderstandinggeneration}, Kimi-Audio~\citep{ding2025kimi}, and MiniCPM-o 2.6~\citep{yao2024minicpm}, significantly expanding the boundaries of human-computer interaction. With the rapid iteration of model capabilities, the challenge of objectively and systematically evaluating these models has become a focal point of academic interest. However, current audio evaluation lacks a unified framework, with datasets and code scattered across various sources, greatly hindering fair and efficient comparisons between models.

In addition to the issue of fragmented evaluation frameworks, the existing evaluation systems also exhibit significant limitations in terms of evaluation depth and language coverage. 
Traditional evaluation tools are often designed for specific tasks, such as automatic speech recognition (ASR) or automatic speech translation (AST), making it difficult to adapt them to audio foundation models (AFMs) with general-purpose interactive capabilities. 
This is particularly true in areas like prompt management and inference parameter tuning, both of which are critical for the comprehensive assessment of AFMs. As model capabilities rapidly evolve, user-centered speech benchmarks are emerging in addition to traditional ASR and AST tasks. However, these benchmarks still face the following core challenges:

(1) Audio codecs, which serve as the backbone of AFMs, lack systematic performance metrics. A codec consists of an audio tokenizer (which converts audio into discrete tokens) and a vocoder (which reconstructs audio from generated tokens). The design of audio codecs directly affects the fidelity and efficiency of the audio representation, which significantly impacts the overall performance of AFMs~\citep{ye2025codec}. Existing methods are broad and provide limited insight into specific performance dimensions.

(2) Current benchmarks are heavily reliant on English, making it challenging to objectively assess models' performance on Chinese. Mainstream benchmarks, such as SpeechTriviaQA~\citep{kyutai2024moshi}, SpeechWebQuestions~\citep{nachmani2023spoken}, and SpeechAlpacaEval~\citep{fang2025llamaomniseamlessspeechinteraction}, are primarily English-centric, leading to inadequate measurement of models' knowledge and language proficiency in the Chinese context.

To address these challenges, we propose \textbf{UltraEval-Audio}, a unified evaluation framework for audio foundation models.
By decoupling data loading, prompt management, inference parameter control, and diverse post-processing and aggregation methods, this framework provides researchers with a unified and flexible evaluation environment. Users can quickly initiate the evaluation process using a simple ``one-click'' automated script. 
This decoupled framework design not only enhances the reproducibility of experiments but also facilitates rapid adaptation and agile extension for researchers.
Furthermore, {UltraEval-Audio} innovatively introduces an audio codec evaluation scheme and several Chinese-language evaluation benchmarks, filling the gaps from both model components and evaluation benchmarks.
Through this full-stack integration, {UltraEval-Audio} aims to standardize and enhance the transparency the entire evaluation process. By providing real-time public audio leaderboards, it strives to advance the field of audio foundation models towards greater interpretability and fairness.
Our contributions are summarized as follows:
\begin{itemize}[leftmargin=*]

\item \textbf{The first unified audio evaluation framework}: UltraEval-Audio supports a wide range of input-output modalities, including ``Text → Audio'', ``Text + Audio → Text'', ``Audio → Text'', and ``Text + Audio → Audio''. The framework supports 10 languages, 14 core task categories and deeply integrates 24 mainstream models and 36 authoritative benchmarks, covering three key areas: speech, environmental sound, and music. With its user-friendly design, the framework offers a ``one-click'' evaluation feature and publicly available leaderboards for transparent comparisons.
\item \textbf{A multi-dimensional evaluation scheme for audio codecs}: We have established a systematic evaluation scheme that covers semantic accuracy, timbre fidelity, and acoustic quality, addressing the lack of widely accepted and systematic multi-dimensional performance metrics for audio codecs.
\item \textbf{Two new Chinese evaluation benchmarks}: We propose \textbf{SpeechCMMLU} and \textbf{SpeechHSK}, which are designed to systematically measure the knowledge proficiency and language fluency of AFMs in the Chinese context.
\end{itemize}

\vspace{-1.0em}
\section{Related Work}
\vspace{-1.0em}
%评测框架的发展对大模型的发展至关重要，这些框架不仅帮助评测出大模型的能力范围，限制和风险并且指导了在实际开发应用中的开发者，例如，最初大模型的主要表现为世界知识和代码，人们关注在MMLU，HumanEval这些数据集的表现，在大模型的表现超过人类后，大模型发展方向转为数学， Function call，人们开始关注MATH500, AIME, BFCL, Tau-bench这些数据集的评测结果
% The development of evaluation frameworks is critical to the advancement of LLMs. These frameworks not only delineate the capabilities, limitations, and potential risks of LLMs but also offer essential guidance for developers in practical application.
In this section, we introduce the latest developments in audio foundation models, audio evaluation frameworks, evaluation benchmarks, and evaluation of audio codecs.

%音频大模型的发展
\textbf{The Advancements of Audio Foundation Models.} 
%在早期阶段，典型的音频大模型是一个简单的ASR+LLM+TTS的管道结构，明显损失了音频中声学信息。当前音频大模型的标准范式逐渐浮现，这个框架由三部分组成：1. 音频tokenizer: 转化原始音频到离散tokens, 保留音频的语音和声学信息
% 
% In the early stage, typical audio foundation models adopted a simple \textbf{ASR+LLM+TTS} pipeline, which inevitably resulted in the loss of important acoustic information. 
% In early audio applications, tasks were typically handled using a pipeline composed of three specialized models: an ASR model, a language model, and a TTS model. This pipeline inevitably leads to the error accumulation of acoustic information.
Emerging audio foundation models typically adopt an \textbf{Audio Codec+LLM} architecture, which generally comprises three core components: (1) an \textbf{audio tokenizer}, which converts raw audio signals into discrete tokens while preserving both semantic and acoustic information;
%2. 大模型主干，负责自回归的token预测和上下文建模 3. 声码器，从输出的音频tokens合成自然声音。
(2) an \textbf{LLM backbone}, responsible for contextual modeling and autoregressive token prediction; and (3) a \textbf{vocoder}, which synthesizes natural speech waveforms from the generated audio tokens.
% 并非所有语音大模型都包含上面所有的组件，基于是否集成vocoder，我们把音频大模型分为两个类别：1）语音理解的大模型：输入文本和语音，输出只有文本(像Qwen-Audio, Gemini1.5) 2)音频生成大模型，输入语音和文本同时出处语音和文本(像GPT-4o-Realtime, Moshi, MiniCPM-o 2.6, Qwen2.5-Omni, Kimi-Audio)
Based on whether they incorporate a vocoder, audio foundation models can be classified into two categories:
(1) audio understanding foundation models, which accept both audio and text as input and produce only text as output (e.g., Qwen-Audio~\citep{chu2023qwenaudioadvancinguniversalaudio,chu2024qwen2audiotechnicalreport}, Gemini-1.5~\citep{geminiteam2024gemini15unlockingmultimodal}). (2) audio generation foundation models, which accept both audio and text as input and generate both speech and text as output (e.g., GPT-4o-Realtime~\citep{openai2024gpt4ocard}, Moshi~\citep{kyutai2024moshi}, MiniCPM-o 2.6~\citep{yao2024minicpm}, Qwen2.5-Omni~\citep{xu2025qwen2}, and Kimi-Audio~\citep{ding2025kimi}). 

Meanwhile, audio codecs are also rapidly evolving. SoundStream~\citep{zeghidour2021soundstream} is the first universal audio codec capable of handling diverse audio types. EnCodec~\citep{défossez2022high}, DAC (descript-audio-codec)~\citep{kumar2023high}, HiFi-Codec~\citep{yang2023hifi}, X-codec~\citep{ye2025codec}, BigCodec~\citep{xin2024bigcodec}, and BiCodec~\citep{wang2025sparktts} further improve reconstruction quality, codebook efficiency, and compatibility with LLM-based speech generation, reflecting a clear trend toward scalable, low-latency, and generative audio tokenizers.

%多模态大模型评测框架
%大模型发展与评测框架发展息息相关，在文本大模型领域，有多个评测框架：HELM， FlagEval， OpenCampss，OpenAI Evals，UltraEval等。在视觉大模型领域，LVLMeHub，VLMEvalKit， HEIM等。然而，尽管音频大模型的流行和一系列的模型发布，还是缺乏一个对于这些模型的全面评测。
% \textbf{Multimodal LLM Evaluation Framework.} 
% The development of LLMs has been closely linked to the evolution of evaluation frameworks. For text-based LLMs, several evaluation frameworks exist: HELM~\citep{liang2022holistic}, FlagEval~\footnote{https://flageval.baai.ac.cn}, OpenCompass~\citep{2023opencompass}, OpenAI Evals~\footnote{https://github.com/openai/evals}, and UltraEval~\citep{he2024ultraeval}. In the domain of visual LLMs, frameworks such as LVLMeHub~\citep{xu2024lvlm}, VLMEvalKit~\citep{duan2024vlmevalkit}, and HEIM~\citep{lee2023holistic} have been proposed. However, despite the rising popularity of audio foundation models and the growing number of released models, a comprehensive evaluation of these models has been lacking.

% 语音评测的发展
\textbf{The Development of Audio Evaluation Frameworks.}
%在音频大模型出现之前，音频领域的任务主要以ASR，AST，TTS为主，其他的比如，情绪识别，声音分类等。每个模型都是为指定的任务而构造，对应的评测也是临时的，附带在模型仓库。比如：ASR的评测主要使用whisper模型代码仓库附带的评测脚本，或者ESPnet-SE
Many comprehensive frameworks have been proposed for evaluating textual and visual foundation models such as OpenCompass~\citep{2023opencompass}, OpenAI Evals~\footnote{https://github.com/openai/evals}, UltraEval~\citep{he2024ultraeval}, and VLMEvalKit~\citep{duan2024vlmevalkit}. However,  a comprehensive evaluation framework for audio foundation models has been lacking.

Before the emergence of audio foundation models, audio models were usually designed for specific tasks, with their evaluation typically being ad hoc and often included alongside the model repository. % For example, ASR evaluations were commonly conducted using scripts from Whisper or ESPnet-SE.
% LLMs是任务无关学习器所以催进了综合评测框架的出现。类似的，音频大模型合并了ASR, AST, TTS, 情绪识别，音色判断等任务。该领域需要一个全面评测音频的框架。AudioBench 为评测音频大模型收集了8个任务26个数据集，但是缺少音频生成的任务。
% LLMs are task-agnostic learners~\citep{NEURIPS2020_1457c0d6}, which has in turn driven the development of comprehensive evaluation frameworks~\citep{liang2022holistic}. 
% Similarly, audio foundation models integrate a wide range of traditional audio tasks, including ASR, AST, TTS, emotion recognition, and others, underscoring the need for a unified and comprehensive evaluation framework in this domain. 
Audio foundation models have demonstrated strong general capabilities across various tasks, making it increasingly necessary to develop a comprehensive evaluation framework that integrates multiple tasks.
Several audio evaluation frameworks have been proposed.
For instance, {AudioBench}~\citep{chen2024voicebench} collects 8 distinct tasks and 26 benchmarks for evaluating audio foundation models. {AHELM}~\citep{lee2025ahelm} aggregates various datasets to holistically measure the performance of AFMs across 10 aspects. But they lack coverage of audio generation tasks. {Kimi-Audio-Evalkit}~\citep{ding2025kimi} integrates all benchmarks mentioned in Kimi-Audio evaluation for reproduction. However, its evaluation process requires five steps, making it cumbersome to use. Additionally, modifying prompts is inconvenient, as changes must be made directly in the code rather than through configuration files. {AU-Harness}~\citep{surapaneni2025harness} offers an efficient evaluation engine supporting over 380 tasks, but it requires users to manually adapt open-source audio foundation models into standardized vLLM services. 
% \textbf{AHELM}~\citep{lee2025ahelm} aggregates various datasets to holistically measure the performance of AFMs across 10 aspects: {audio perception, knowledge, reasoning, emotion detection, bias, fairness, multilinguality, robustness, toxicity, and safety}. However, AHELM shares the same code repository as HELM, which makes installation, configuration, and usage not straightforward.

%与此同时，不同于传统的ASR, AST数据集，领域内开始逐渐建立面向用户使用为中心的数据集，这些数据集以纯语音形式输入，评测模型的直接回复。在音频理解模态，AIR-Bench 从现有的数据集中收集了语音QA并且使用GPT-4来评测模型回复。Voicebench收集了一些音频QA同时也从文本指令数据集(AlpacaEval,IFEval)中通过Google-TTS合成了音频。 
\textbf{The Development of Audio Evaluation Benchmarks.}
Beyond traditional ASR and AST benchmarks, the field has begun developing user-centric benchmarks that use raw speech as input without additional task description and directly evaluate model responses. For audio understanding, AIR-Bench~\citep{yang2024air} collects spoken question answering (QA) samples from existing datasets and employs GPT-4 as an automatic evaluator. VoiceBench~\citep{chen2024voicebench} further expands this direction by including both naturally spoken QA samples and synthetic spoken instructions, which are generated from text-based instruction-following datasets (e.g. AlpacaEval~\citep{alpaca_eval}, IFEval~\citep{zhou2023instructionfollowingevaluationlargelanguage}) using Google TTS. 
%在音频生成模态，Llama-Question是第一个合成语音QA数据集，并且给出评测范式：通过conformerASR转写回复音频来评测准确率。Sepeech WebQuestion 继承WebQuestion, Speech TrivaQA同样合成音频来自TriviaQA. Sepeech AlpacaEval选择了适合音频交互场景的数据从AlpacaEval。 然而上面这些数据都是英文，缺少其他它语言
For audio generation, the first dedicated speech question-answering benchmark, Llama-Question~\citep{nachmani2023spoken}, introduced a synthetic speech QA dataset with a novel evaluation paradigm: it employs ConformerASR~\citep{gulati2020conformer} to transcribe reply audio into text before assessing answer accuracy. SpeechWebQuestions~\citep{nachmani2023spoken}, SpeechTriviaQA~\citep{kyutai2024moshi}, and SpeechAlpacaEval~\citep{fang2025llamaomniseamlessspeechinteraction} are derived from corresponding textual benchmarks WebQuestions~\citep{chen2015microsoft}, TriviaQA~\citep{joshi2017triviaqa}, and AlpacaEval.
% SpeechWebQuestions~\citep{nachmani2023spoken} is derived from WebQuestions~\citep{chen2015microsoft}, while SpeechTriviaQA~\citep{kyutai2024moshi} similarly synthesizes audio from the TriviaQA~\citep{joshi2017triviaqa} dataset. SpeechAlpacaEval~\citep{fang2025llamaomniseamlessspeechinteraction} selects suitable data for speech interaction scenarios from AlpacaEval. 
However, all these benchmarks are currently limited to English, leaving multilingual speech benchmarks largely unexplored.

\textbf{The Development of Audio Codec Evaluation.} The evaluation of audio codecs employs both subjective and objective metrics. Subjective evaluation typically follows the MUSHRA~\citep{series2014method} protocol, which uses both a hidden reference and a low anchor. Objective evaluation includes several approaches: ViSQOL~\citep{hines2015visqol,chinen2020visqol} measures spectral similarity to the ground truth as a proxy for mean opinion score; Scale-Invariant Signal-to-Noise Ratio (SI-SNR) quantifies the similarity between reconstructed and original audio while ignoring signal scale; Mel distance computes the difference between the log-Mel spectrograms of reconstructed and ground truth waveforms; STOI~\citep{taal2011algorithm} assesses speech intelligibility; and speaker similarity (SIM) is calculated as the cosine similarity between speaker vectors of the reconstructed audio and ground truth using an embedding model. Beyond these direct metrics, recent works like that of~\cite{ye2025codec} also employ downstream tasks such as TTS to indirectly evaluate codec performance.

\vspace{-1.0em}
\section{Audio Evaluation Design and Methodology}
\vspace{-1.0em}

%本章旨在构建一套系统化的音频基础模型评测体系和方法论。首先， 3.1节提出了一种音频评测的全面体系，该体系任务分类和基准实例化。在此基础上，考虑到音频编解码器在音频基础模型架构中的核心作用，3.2节提出音频编解码器全维度评估方法。针对于当前研究中中文评估资源的匮乏，3.3 节将详细介绍我们构建的两个中文语音评估基准——SpeechCMMLU与SpeechHSK。
This section outlines a systematic audio evaluation design and methodology  for audio foundation models.
Section~\ref{sec:Audio_Evaluation_Taxonomy} introduces a unified audio evaluation taxonomy that organizes tasks and their corresponding benchmark instantiations. 
Building upon this taxonomy, Section~\ref{sec:codec_evaluation} focuses on codec evaluation and develops a comprehensive methodology for assessing audio codecs, a core component of audio foundation model architectures.
Finally, Section~\ref{sec:chinese_benchmarks} introduces two Chinese speech benchmarks, SpeechCMMLU and SpeechHSK, to address the lack of systematic evaluation resources for Chinese in existing audio evaluation research.

\vspace{-1.0em}
\subsection{Audio Evaluation Taxonomy}
\vspace{-1.0em}
\label{sec:Audio_Evaluation_Taxonomy}
%当前音频AI领域模型在能力形态上呈现出显著的多样性：不同输入输出模态、多种任务类型、以及跨语言与跨领域的泛化需求。现有很多评测工作多围绕单一任务或单一模态构建，缺乏一个统一的、可扩展的评测组织方式。UltraEval-Audio 并非简单整合 benchmarks，而是构建一套可复用、可对齐、可扩展的音频评测分类体系，为后续评测设计给出方论指引。
Audio foundation models cover a wide range of modalities, tasks, languages and domains. Existing evaluation efforts typically focus on isolated tasks or specific modalities, underscoring the need for a unified and extensible evaluation framework. Rather than simply aggregating existing benchmarks, UltraEval-Audio introduces a reusable and extensible evaluation taxonomy that offers methodological guidance for future evaluation design.

\textbf{Task taxonomy: What Capabilities Are Evaluated}

%UltraEval-Audio采用一种面向能力的任务分类体系，以系统性地刻画音频基础模型的评估空间。将常见的评估目标抽象为一组核心任务类别，每个类别对应音频模型应展现的一项特定能力。该分类体系在设计上兼顾能力目标与输出模态，使得具有相似语义目标但不同输出形式的任务能够被区分对待，从而更准确地反映模型能力差异。如\ref{tab:Task taxonomy in UltraEval-Audio.}所示，所有任务主要围绕三大类别：音频理解，音频生成和音频编解码，并进一步按语音、音乐、环境声等应用领域细分。
UltraEval-Audio adopts a capability-driven task taxonomy to organize the evaluation of audio foundation models.
Evaluation tasks are grouped into core categories that reflect distinct model capabilities and typical input–output settings.
As summarized in Table~\ref{tab:task_overview_en}, the taxonomy consists of three high-level categories: audio understanding, audio generation, and audio codec.
Within each category, tasks are further organized by application domains, including speech, music, and environment sounds.

\begin{table*}[!t]
  \centering
  \small
  \renewcommand{\arraystretch}{1.2}
  \caption{Task taxonomy in UltraEval-Audio.}
  \resizebox{\textwidth}{!}{
  \begin{tabular}{lcclc}
    \toprule
    \textbf{Category} & \textbf{Domain} & \textbf{Task} & \textbf{Description} & \textbf{Metrics} \\
    \midrule
    \multirow{10}{*}{\shortstack[l]{Audio\\Understanding}} 
      & \multirow{5}{*}{Speech} 
        & ASR & Given speech audio, produce a transcription. & WER / CER \\
      & & AST & Given speech audio in the source language, generate a text translation in the target language. & BLEU \\
      & & Gender Analysis & Given speech audio, predict the speaker’s gender. & Acc. \\
      & & Speech QA & Given speech audio, generate a textual answer to the corresponding question. & Exist Match / G-Eval \\
      & & Emotion Analysis & Given speech audio, identify the speaker’s emotional state. & Acc. \\
    \cmidrule{2-5}
      & \multirow{3}{*}{Music} 
        & Instrument Recognition & Given music audio, classify the predominant instrument. & Acc. \\
      & & Music Genre & Given music audio, classify the corresponding music genre. & Acc. \\
      & & Chord Recognition & Given music audio, identify the sequence of chord labels. & Acc. \\
    \cmidrule{2-5}
      & \multirow{2}{*}{\shortstack[l]{Environment\\Sounds}} 
        & Audio Classification & Given non-speech audio, classify it into a predefined scene or event category. & Acc. \\
      & & Audio Captioning & Given non-speech audio, generate a natural language description of its content. & BLEU / ROUGE-L \\
    \midrule
    \multirow{3}{*}{\shortstack[l]{Audio\\Generation}} 
      & \multirow{3}{*}{Speech} 
        & TTS & Given input text, synthesize the corresponding speech audio. & ASR-WER \\
      & & VC & Given input text and a reference speech sample, synthesize speech in the target speaker’s voice. & ASR-WER / SIM \\
      & & Speech QA & Given speech audio, generate an appropriate spoken response. & Exist Match / G-Eval / UTMOS \\
    \midrule
    Audio Codec & Speech & Speech Codec & Encode and reconstruct speech audio from discrete representations. & ASR-WER / SIM / UTMOS \\
    \bottomrule
  \end{tabular}
  }

  \vspace{4pt}
  \footnotesize
  \raggedright
  \textbf{Notes: }
  WER: Word Error Rate; CER: Character Error Rate; BLEU/ROUGE-L: text generation quality; 
  Acc.: classification accuracy; SIM: speaker embedding cosine similarity; 
  UTMOS: An objective speech quality evaluation metric; 
  G-Eval: GPT-based evaluation metric; 
  ASR-WER: computed by transcribing the generated or reconstructed speech with an ASR model and then calculating WER on the transcriptions.
  \label{tab:task_overview_en}
\end{table*}
%音频理解任务侧重于评估模型感知、识别以及从语义上解析音频信号的能力。在语音领域，这包括自动语音识别（ASR）、语音翻译（AST）、说话人相关属性分析（如性别识别）以及语气、情感、语速等理解任务。除识别与分类外，通过语音问答任务可以评估更高层级的语义推理能力，此类任务要求模型理解口语内容并生成基于文本的回答。对于非语音音频，音频理解任务涵盖音乐相关的感知能力，包括乐器识别、音乐流派分类与和弦识别；同时也包括通过音频分类与音频描述任务实现的环境声理解，其中音频描述任务要求模型对声学场景或事件生成自由形式的文本描述。
Audio understanding tasks evaluate a model’s ability to extract and interpret semantic information from audio inputs. This category spans multiple levels of analysis across speech and non-speech domains.
In the speech domain, tasks range from low-level recognition such as ASR and AST to higher-level semantic reasoning such as speech QA.
In the non-speech domain, tasks include low-level classification like instrument recognition and higher-level comprehension such as audio captioning.
% where the latter requires generating free-form textual descriptions of acoustic scenes or events.

%音频生成任务评估模型在给定条件下合成或转换音频的能力。在语音领域，这包括文本到语音合成（TTS）、语音克隆（VC）以及面向语音问答的口语回答生成。尽管语音问答在语义目标上与音频理解任务相似，但其生成语音输出的形式使其增加了考察模型在语音生成质量、自然度与可懂度等方面的能力。
Audio generation tasks assess a model’s ability to synthesize or transform audio under given conditions. In the speech domain, tasks include TTS, voice clone (VC), and spoken answer generation for speech QA tasks. In audio understanding tasks, Speech QA evaluates the accuracy of textual responses.
In audio generation tasks, it additionally assesses the quality of generated audio responses in terms of acoustics, naturalness, and intelligibility.

%此外，我们将音频编解码评估作为一类独立的评测任务进行建模。尽管音频编解码并非传统意义上的应用任务，但其在音频基础模型中承担着中间表示与信息压缩的核心功能，对下游理解与生成能力具有重要影响。因此，通过语音编解码任务可以系统性地评估模型在保持语义与声学特征前提下进行音频压缩与重建的能力。
Additionally, audio codec evaluation is treated as a separate task category. Although audio codecs are not traditional application tasks, they play an important role in audio foundation models. Audio codec tasks systematically assess a codec’s ability to compress and reconstruct audio while preserving both semantic content and acoustic characteristics.

%每个任务均与其能力目标相对应的标准化评估指标关联。例如，语音识别采用词错误率/字符错误率，翻译与描述任务采用BLEU和ROUGE等文本相似度指标，属性与声音识别任务采用分类准确率，音频生成与编解码评估则采用基于自动语音识别的WER或语音质量指标。上述任务分类体系为系统组织多样化音频评测基准提供了统一且可扩展的基础，从而保证了对不同音频领域与模型类型能力覆盖的一致性与可比性
Each task category is associated with standardized evaluation metrics aligned with its target capabilities.
For example, ASR uses word error rate (WER) or character error rate (CER); translation and captioning tasks rely on text similarity metrics such as BLEU and ROUGE; attribute and sound classification tasks use accuracy; and audio generation and codec tasks are evaluated with ASR-based WER or dedicated speech quality metrics.

By organizing tasks and metrics in this way, UltraEval-Audio provides a reusable and extensible evaluation taxonomy that offers methodological guidance for future evaluation design, ensuring consistent and comparable coverage of capabilities across audio domains and model types.

\textbf{Benchmark Instantiation: How Capabilities Are Measured}

\begin{table*}[!t]
  \centering
  \small
  \renewcommand{\arraystretch}{1.2}
  \caption{Benchmarks supported in UltraEval-Audio. * indicates new benchmarks introduced in this paper.}
  \resizebox{\textwidth}{!}{
  \begin{tabular}{lcc}
    \toprule
    \textbf{Task}  & \textbf{Language} & \textbf{Dataset}\\
    \midrule

\multirow{12}{*}{ASR}
  & \multirow{3}{*}{en} & TED-LIUM~\citep{rousseau2012ted}, VoxPopuli~\citep{wang2021voxpopuli}, LibriSpeech~\citep{panayotov2015librispeech}\\
  &                      & The People's Speech~\citep{galvez2021people}, WenetSpeech~\citep{zhang2022wenetspeech} \\
  &                      & GigaSpeech~\citep{chen2021gigaspeech}, AudioMNIST~\citep{sripaadsrinivasan2023audiomnist} \\
  \cmidrule(lr){2-3}
  & \multirow{1}{*}{zh}  & KeSpeech~\citep{tang2021kespeech}, AISHELL-1~\citep{bu2017aishell} \\
  \cmidrule(lr){2-3}
  & nl, fr, de, it, pl, pt, es & MLS~\citep{Pratap2020MLSAL} \\
  \cmidrule(lr){2-3}
  & \multirow{1}{*}{zh, en, ru, de, jp, ...} & FLEURS~\citep{fleurs2022arxiv}, Common Voice~\citep{ardila2019common} \\
\midrule

AST & zh, en, ru, de, jp, ... & CoVoST 2~\citep{wang2020covost} \\
\midrule

\multirow{1}{*}{VC}
  & \multirow{1}{*}{zh, en} & Seed-TTS-Eval~\citep{anastassiou2024seed}, CV3-Eval~\citep{du2025cosyvoice} \\
\midrule

TTS & zh, en & Long-TTS-Eval~\citep{wang2025mgm} \\
\midrule

\multirow{2}{*}{Speech Codec}
  & en & LibriSpeech \\
  \cmidrule(lr){2-3}
  & zh & AISHELL-1 \\
\midrule

\multirow{3}{*}{Speech QA}
  & \multirow{3}{*}{en} & SpeechTriviaQA~\citep{kyutai2024moshi}, SpeechWebQuestions~\citep{nachmani2023spoken} \\
  &                     & SpeechAlpacaEval~\citep{fang2025llamaomniseamlessspeechinteraction}, LLaMA-Questions~\citep{nachmani2023spoken} \\
  &                     & AIR-Bench~\citep{yang2024air}, MMAU~\citep{sakshi2024mmau} \\
  \cmidrule(lr){2-3}
  & \multirow{1}{*}{zh} & SpeechHSK\textsuperscript{*}, SpeechCMMLU\textsuperscript{*} \\
\midrule

\multirow{1}{*}{Emotion Analysis}
  & \multirow{1}{*}{en} & TESS~\citep{dupuis2010toronto}, MELD~\citep{poria2018meld} \\
\midrule

Gender Analysis & en & VoxCeleb~\citep{nagrani2017voxceleb} \\
\midrule

Chord Recognition & - & Chord~\citep{deepcontractor2023chord} \\
\midrule

Instrument Recognition & - & NSynth~\citep{engel2017neural} \\
\midrule

Music Genre & - & GTZAN~\citep{sturm2013gtzan} \\
\midrule

\multirow{1}{*}{Caption}
  & \multirow{1}{*}{-} & AudioCaps~\citep{kim2019audiocaps}, WavCaps~\citep{mei2024wavcaps}, Clotho~\citep{drossos2020clotho} \\
\midrule

\multirow{3}{*}{Audio Classification}
  & \multirow{3}{*}{-} & CatDog~\citep{mmoreaux2023catsdogs}, DESED~\citep{turpault2019sound} \\
  &                    & VocalSound~\citep{gong_vocalsound}, COVID-19 Sounds~\citep{dong2020interactive} \\
  &                    & PASCAL CHSC 2011~\citep{pascal-chsc-2011}, ICBHI 2017 Respiratory Sound~\citep{rocha2018alpha}  \\
    \bottomrule
  \end{tabular}
  }
  \label{tab:supported_benchmarks}
\end{table*}

%基于上述任务分类体系，UltraEval-Audio通过系统性挑选一系列具有代表性且被广泛采用的基准测试，对每项评估能力进行了实例化。如表~\ref{tab:supported_benchmarks} 所总结，该框架目前已整合总计36个基准测试，覆盖广泛的任务和语言。每个基准测试都明确映射到预定义的任务类别，确保了能力定义与具体评估方案之间的一致性。
Building on the task taxonomies, UltraEval-Audio instantiates each evaluated capability through a carefully selected set of widely adopted benchmarks. As summarized in Table~\ref{tab:supported_benchmarks}, the framework integrates 36 benchmarks covering 14 tasks and 10 languages. Each benchmark is explicitly mapped to a predefined task category, ensuring consistency between capability definitions and specific evaluation details.

%每个任务类别下，我们都汇集了多个基准数据集，以实现对模型能力的多维度、鲁棒性评估。以音频理解中的自动语音识别（ASR）任务为例，我们支持包括 LibriSpeech、Common Voice、AISHELL-1、WenetSpeech 在内的十多个数据集，涵盖了从清晰朗读语音（如 LibriSpeech）到复杂嘈杂场景（如 WenetSpeech）、从单一语言（如 AISHELL-1 之于中文）到多语言（如 MLS、FLEURS）的广泛测试条件。这种设计允许我们不仅评估模型的绝对识别准确率（WER/CER），还能评估其在不同口音、领域、噪音水平和语言上的泛化能力与鲁棒性。类似地，对于语音问答（Speech QA），我们整合了 SpeechTriviaQA、SpeechWebQuestions 等多个知识型和指令遵循型基准。
Within each task, multiple benchmark datasets are aggregated to enable multidimensional and robust assessment of model capabilities. For the ASR task, we support over ten datasets. These datasets span clear read speech (e.g., LibriSpeech) to complex noisy scenarios (e.g., WenetSpeech), and cover single-language (e.g., AISHELL-1 for Chinese) as well as multilingual conditions (e.g., MLS, FLEURS). This design allows evaluation not only of absolute recognition accuracy (WER/CER) but also of models’ generalization and robustness across accents, domains, noise levels, and languages. Similarly, for Speech QA, we incorporate multiple knowledge-based and instruction-based benchmarks, including SpeechTriviaQA and SpeechAlpacaEval.

%在所有任务中，UltraEval-Audio的基准测试横跨多种语言和声学领域，并使用任务分类体系中定义的、适用于各任务的指标进行评估。这一基准测试实例化策略确保了评估框架既全面又可扩展，同时在抽象能力定义与具体测量流程之间保持了清晰一致的对齐关系。
Across all tasks, the selected benchmarks cover multiple languages and diverse acoustic domains, and are evaluated using metrics specified in the task taxonomy and tailored to each task. This instantiation strategy ensures that the evaluation framework is comprehensive, extensible, and tightly aligned with the underlying capability definitions, providing a clear foundation for systematic and reproducible assessment.

\vspace{-1.0em}
\subsection{Codec Evaluation}
\vspace{-1.0em}
\label{sec:codec_evaluation}
%作为音频模型的底层基础组件，音频编解码器的设计直接决定了音频表征的保真度与处理效率，进而显著影响上层音频基础模型的整体性能。然而，现有评测工作存在指标分散、标准不统一的问题，缺乏系统化、可比性的评估方法。为全面、客观地衡量编解码器在压缩与重建过程中的综合表现，我们构建了一个涵盖语义、音色保真度与声学质量的三维评测体系。
Audio codecs are a fundamental component of audio foundation models, as their design directly affects the fidelity of audio representations and overall model performance. Existing evaluation approaches, however, rely on diverse metrics with inconsistent standards, and lack a systematic, comparable assessment framework. To address this, we propose a three-dimensional codec evaluation methodology encompassing \textbf{semantics}, \textbf{timbre fidelity}, and \textbf{acoustic quality}.

%在语义方面，我们采用词错误率（WER）量化重建音频对原始内容的保留程度。具体做法是通过高性能自动语音识别系统将重建音频转写为文本，并与原文比对，英文测试采用 Whisper-large-v3 模型，中文测试使用针对中文优化的 Paraformer-zh 模型。在音色保真度方面，我们使用在说话人验证任务上精调的 WavLM-large 模型提取音频嵌入，并计算原始与重建音频嵌入的余弦相似度，从而量化编解码器在保留说话人音色特征上的能力。在声学感知质量方面，为全面评估音频的自然度与听感舒适度，我们结合了 UTMOS 指标预测整体自然度，并辅以 DNSMOS P.835 与 P.808 指标评估在噪声环境下的语音质量表现。
For semantics, we measure how well reconstructed audio preserves the original content using WER. Specifically, the reconstructed audio is transcribed by high-performance ASR models and compared to the original transcript. We employ Whisper-large-v3 for English and Paraformer-zh for Chinese.
For timbre fidelity, we extract audio embeddings using WavLM-large fine-tuned on speaker verification, and compute the cosine similarity between embeddings of the original and reconstructed audio. This quantifies the codec’s ability to retain speaker characteristics.
For acoustic quality, we assess the naturalness and perceptual comfort of audio using a combination of UTMOS~\citep{saeki2022utmosutokyosarulabvoicemoschallenge} to predict overall naturalness, alongside DNSMOS P.835~\citep{reddy2022dnsmos} and P.808~\citep{reddy2021dnsmos} to evaluate speech quality in noisy environments.

%该三维评估框架从内容、说话人特征和听感三个核心层面提供了互补的视角，能够更全面、更可靠地揭示音频编解码器的性能特性与潜在局限，为模型研发与优化提供了精细化的诊断工具。
This three-dimensional evaluation framework provides complementary perspectives across content, speaker characteristics, and perceptual quality, enabling a more comprehensive and reliable characterization of codec performance and potential limitations, and offering a fine-grained diagnostic tool for model development and optimization.

\vspace{-1.0em}
\subsection{Chinese Benchmarks}
\vspace{-1.0em}
\label{sec:chinese_benchmarks}
%音频大语言模型的发展推动语音评测从传统低层指标（如 ASR 词错误率）向复杂语义理解与知识推理等高层能力转型。然而，现有高层次语音评测资源严重偏向英语，例如SpeechTriviaQA, SpeechWebQuestions，SpeechAlpacaEval等，中文语境下基准测试仍相对匮乏。为缓解这个问题，我们推出两项全新基准测试：SpeechCMMLU和SpeechHSK。前者旨在评估模型在中文世界知识任务上的复杂语音理解与推理能力，后者则作为中文语言理解能力评估。
The development of audio foundation models shift speech evaluation from traditional low-level metrics like WER, toward higher-level capabilities including complex semantic understanding and knowledge reasoning. However, existing high-level speech benchmarks are largely focused on English, such as SpeechTriviaQA, SpeechWebQuestions, and SpeechAlpacaEval, leaving a relative scarcity of resources in the Chinese context. To address this gap, we introduce two new benchmarks: \textit{SpeechCMMLU} evaluates models’ ability to comprehend and reason over Chinese world-knowledge tasks, while \textit{SpeechHSK} measures Chinese language proficiency.

%SpeechCMMLU通过系统性语音合成方法将广受认可的中文知识推理评测集CMMLU扩展至语音模态。下面描述了我们的生成和质检方法：
\textbf{SpeechCMMLU} extends the widely recognized Chinese knowledge reasoning benchmark CMMLU~\citep{li2023cmmlu} into the audio modality through systematic speech synthesis. The construction and quality assurance process is as follows:
\begin{enumerate}[leftmargin=*]
% 文本指令构建：将每条选择题组织为统一的指令格式，以便后续语音生成与解析。具体形式为将题干、选项及作答约束集成为标准化指令.
\item \textbf{Text instruction construction}. Each multiple-choice question is formatted as a unified instruction, combining the question stem, options, and answer constraints into a standardized prompt.
% 音频指令合成：考虑到人工录制成本高昂且数据规模大（11583 条），采用高质量 TTS 模型 CosyVoice2 生成语音指令。该模型在中文发音清晰度及专业术语表现方面表现稳定，适合大规模语音数据构建。
\item \textbf{Audio instruction synthesis}. Considering the high cost of manual recording and the large scale of the dataset (11,583 items), we employ the high-quality TTS model CosyVoice2 to generate the speech prompts. This model demonstrates consistent performance in Chinese pronunciation and specialized terminology, making it suitable for large-scale speech dataset creation.
% 自动化质检：为了保证语义保真性，对所有合成音频进行 ASR 转写。仅当转写文本与原始文本完全一致（字符错误率 CER=0%）的样本被保留，从而有效排除 TTS 合成过程中可能引入的发音误差或术语偏差。
\item \textbf{Automated quality control}. To ensure semantic fidelity, all synthesized audio is transcribed via a ASR process. Only samples whose transcription exactly matches the original text (CER = 0\%) are retained, effectively eliminating potential pronunciation errors or terminology deviations introduced during TTS synthesis.
\end{enumerate}
%经过上述处理，最终发布的SpeechCMMLU包含3,519条高质量语音样本，覆盖广泛学科主题，适用于评测模型在专业领域的中文知识理解与推理能力。
Following this procedure, the released SpeechCMMLU dataset contains 3,519 high-quality speech samples spanning diverse academic subjects, suitable for evaluating models’ Chinese knowledge comprehension and reasoning capabilities in professional domains.

%汉语水平考试（HSK）是中国教育部设立的全球性中文能力标准化测评体系，其权威性与科学性已获国际广泛认可。SpeechHSK将HSK听力理解部分转化为语音评测基准，为模型的中文语言技能提供层级化、实用化的评估标尺。
\textbf{SpeechHSK} leverages the Hanyu Shuiping Kaoshi (HSK), the standardized global Chinese proficiency test established by the Chinese Ministry of Education, widely recognized for its authority and scientific rigor. SpeechHSK converts the listening comprehension portion of HSK into an audio benchmark, providing a hierarchical and practical measure of models’ Chinese language skills.

\begin{itemize}[leftmargin=*]
%- 数据来源与构建：基准数据来源于HSK官方真题的听力理解部分。每条样本由题干及四个选项组成，其中题干音频使用原始考试录音，而选项原本为文本形式，为适配语音评测，我们由中文母语者将文本选项重新录制为音频，确保了语音自然度。
    \item \textbf{Data sources and construction}. The benchmark draws from official HSK listening comprehension questions. Each sample consists of a question and four options, where the question audio uses the original exam recordings, while the options are originally provided in text form. To adapt textual options for speech evaluation, native Chinese speakers re-record them as audio, ensuring naturalness of the audio.
%- 难度分级结构：数据集严格遵循HSK的六个等级（1级为入门，6级为精通）进行组织，题目数量随等级提升而增加，共计170条样本。这种结构支持对模型中文理解能力进行渐进式、诊断性评估。
    \item \textbf{Difficulty-level structure}. The dataset follows HSK’s six proficiency levels (Level 1: beginner; Level 6: advanced), with an increasing number of questions at higher levels, totaling 170 samples. This structure supports progressive, diagnostic assessment of models’ Chinese language comprehension.
\end{itemize}

Together, SpeechCMMLU and SpeechHSK form a multi-level evaluation suite for Chinese speech. SpeechCMMLU focuses on knowledge reasoning, while SpeechHSK emphasizes general language proficiency. Both benchmarks are constructed via automated pipelines with rigorous quality control, ensuring reproducibility and scalability.

\vspace{-1.0em}
\section{UltraEval-Audio Framework}
\vspace{-1.0em}

%本章详细介绍了 UltraEval-Audio 的工程架构设计。为了解决音频基础模型（AFMs）评测中因模态多样与接口多样导致的集成难题，本框架采用了高度解耦的模块化设计（Decoupled Modular Design）与配置驱动（Configuration-driven）的工作流。这种设计使得各功能组件可以独立开发、灵活扩展与高效复用。通过统一的配置文件（YAML）进行数据、模型与评测流程的声明式管理，用户无需手动编写脚本即可实现“一键式”自动化评测。
This section presents the engineering architecture of UltraEval-Audio. To address the integration challenges in the evaluation of audio foundation models arising from diversity in input modalities and model interfaces, the framework adopts a decoupled modular design together with a configuration-driven workflow. This design enables components to be developed independently, extended flexibly, and reused efficiently. All datasets, models, and evaluation pipelines are declaratively specified through unified YAML configuration files, enabling end-to-end evaluation without manual scripting.

\begin{figure}[htb]
  \centering
  \includegraphics[width=\columnwidth]{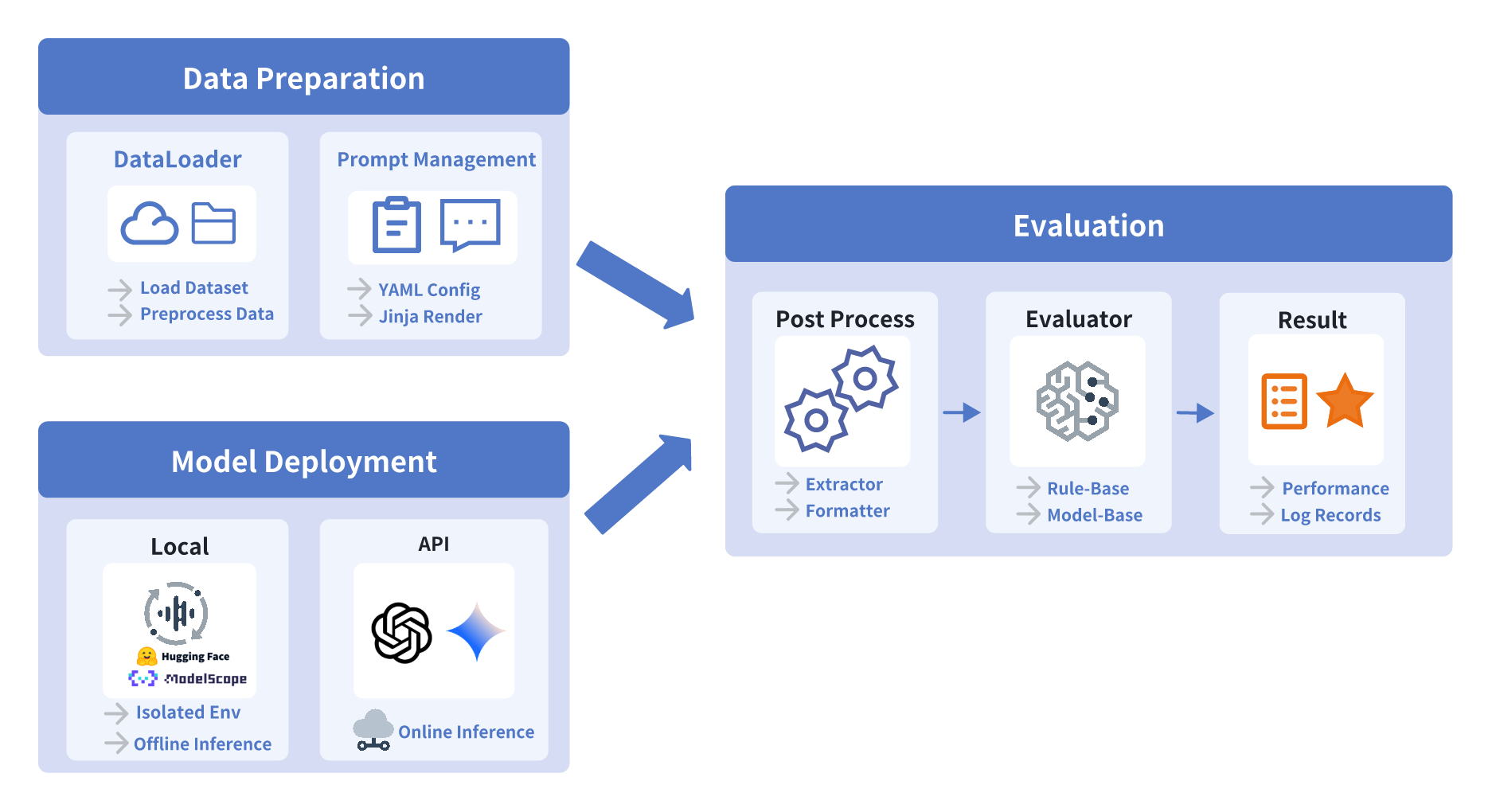}
  \caption{Overview of the UltraEval-Audio framework}
  \label{fig:runtime}
\end{figure}

%如图 1 所示，系统由三大阶段构成：数据准备 （Data Preparation）、模型部署 (Model Deployment)和评测 (Evaluation)。接下来的章节将逐一解析这些组件：4.1 节描述了多模态数据导入与提示词管理；4.2 节展示了模型部署的创新设计；4.3 节展示了评测执行流程及自动化评估模块的实现细节。
 As illustrated in Figure~\ref{fig:runtime}, the framework consists of three main modules: data preparation, model deployment, and evaluation. The following subsections provide a detailed description of each component. Section~\ref{sec:data_preparation} details data loader and prompt management, section~\ref{sec:model_delpoyment} presents the model deployment mechanisms, and section~\ref{sec:evaluation} describes the evaluation pipeline and automated scoring modules.

\vspace{-1.0em}
\subsection{Data Preparation}
\vspace{-1.0em}
\label{sec:data_preparation}

\textbf{Data loader.}
%在音频评测中，数据的加载与管理是评测系统的基础设施。与结构单一的纯文本数据集不同，音频数据集呈现出显著的多模态耦合特性，通常由非结构化的音频信号文件与结构化的文本标注共同组成。这种特性引入了复杂的数据管理挑战，包括处理分散的文件存储路径、兼容异构的编解码格式以及统一不一致的采样率。
Data loading and management form the foundational infrastructure of audio evaluation. Unlike textual datasets, which typically have simple structures, audio datasets involve a strong coupling between audio signals and textual annotations. This coupling introduces significant data management challenges, including managing dispersed file paths, handling diverse audio encoding formats, and normalizing inconsistent sampling rates.

%为应对上述挑战，UltraEval-Audio 确立了以 音频数据集（Audio-centric Dataset） 为核心的组织范式。在该体系中，特定的 audio 字段被设计用于存储音频内容。框架内置的自动化管线能够在加载阶段动态完成音频资源的获取（支持 HuggingFace 等开源社区的自动下载与缓存）、解码及格式标准化。此外，为了适应多样化的存储环境，Data Loader 亦提供了对本地文件系统数据集的无缝集成与管理支持。
To address these challenges, UltraEval-Audio adopts an audio-centric dataset organization scheme. In this design, a reserved field \textit{audio} is used to store audio content. The framework provides built-in automated pipelines that dynamically manage audio resource acquisition during the loading, including automatic downloading and caching from open-source repositories such as HuggingFace, as well as decoding and format standardization.

%通过上述设计，框架将复杂的多模态预处理解耦为标准化的数据接口，显著简化了准备流程，保障了后续模型推理与评测的高效执行。
The framework decouples complex multimodal preprocessing into a standardized data interface, significantly simplifying the preparation workflow and ensuring efficient execution of subsequent model inference and evaluation.

\textbf{Prompt management.}
%提示词（Prompt） 在引导模型生成任务相关输出方面发挥着关键作用。Prompt 的设计不仅与模型输入形式有关，更直接影响评测结果的稳定性与可比性。然而，不同模型由于训练策略差异，往往对提示词格式（如特殊 Token、角色定义）具有高度敏感性，且在不同任务间差异显著，这种模型敏感性与任务异构性导致硬编码的提示词难以在公平的基准上比较模型性能。
Prompts play a critical role in guiding models to produce relevant task outputs. Prompt design not only relate to the format of model input but also directly affects the stability and comparability of evaluation results. However, due to differences in training strategies, models are often highly sensitive to prompt, such as the use of special tokens or role definitions, and this sensitivity can vary substantially across tasks. The combination of model-specific sensitivity and task-dependent variation makes hard-coded prompts difficult to reuse and undermines fair comparison across models.

%为解决上述问题，UltraEval‑Audio 在框架设计中引入了一个**配置驱动（configuration‑driven）**的 Prompt 管理机制。系统采用基于 YAML 的配置文件描述 Prompt 结构，并结合 Jinja 模板引擎 实现变量注入与条件逻辑控制，从而支持任务级、模型级乃至样本级的灵活提示词构建。通过这种配置化设计，研究者无需修改任何代码，即可快速定义或替换不同任务的 Prompt 模板，全面提升了评测流程的可维护性与可扩展性。
To address this issue, UltraEval-Audio introduces a configuration-driven prompt management mechanism. Prompt structures are defined using YAML-based configurations and combined with the Jinja templating engine to support variable injection and conditional logic. This design enables flexible prompt construction at the task, model, and even sample level. Consequently, researchers can define or replace prompt templates for different tasks without modifying any code, greatly enhancing the maintainability and extensibility of the evaluation workflow.

%在具体实现上，UltraEval‑Audio 的 Prompt 管理模块由三部分组成：
In practice, the prompt management module in UltraEval-Audio comprises three components:
\begin{itemize}[leftmargin=*]
    % Prompt Registry 用于注册并索引所有任务对应的提示模板；
    \item \textbf{Prompt Registry}, which registers and indexes prompt templates associated with different tasks.
    % YAML Parser 负责解析声明式配置文件，加载各模型或任务的 Prompt 定义
    \item \textbf{YAML Parser}, which interprets declarative configuration files and loads prompt definitions for specific models or tasks.
    % Template Renderer 基于 Jinja 渲染机制，将来自数据集的字段（如)注入模板，生成最终输入。
    \item \textbf{Template Renderer}, which leverages Jinja-based rendering to inject dataset fields (e.g., \{\{audio\}\}, \{\{question\}\}, \{\{choice\_a\}\}) into templates and produce the final model inputs.
\end{itemize}
% 该机制不仅支持固定模板，也支持带有动态变量与条件分支的复杂构造。
This mechanism supports not only fixed templates but also more complex prompt constructions involving dynamic variables and conditional branches.
The following is an example of a MiniCPM-o 2.6 ASR prompt:
\begin{verbatim}
mini-cpm-omni-asr-en:
  class: audio_evals.prompt.base.Prompt
  args:
    template:
    - role: user
      contents:
      - type: text
        value: 'Please listen to the audio snippet carefully and transcribe the content. Please 
        output in low case.'
      - type: audio
        value: '{{audio}}'
\end{verbatim}
the \{\{audio\}\} placeholder is dynamically replaced at inference time with the audio field from the corresponding data sample. To adapt the prompt for different ASR models, users only need to define a model-specific prompt and select it at runtime via the \textit{-{}-prompt \$prompt} argument, without modifying any code.

For tasks that require sample-dependent prompt structures, such as multiple-choice questions with a variable number of options, the template can incorporate conditional rendering logic. An example is shown below:

\begin{verbatim}
single_choice_extended:
  class: audio_evals.prompt.base.Prompt
  args:
    template:
      - role: user
        contents:
          - type: audio
            value: "{{audio}}"
          - type: text
            value: "Choose the most suitable answer from options A, B, C, D{% if choice_e is defined 
            and choice_e %}, and E{% endif %} to respond to the question below. 
            You may only choose A, B, C, or D{% if choice_e is defined and choice_e %}, 
            or E{% endif %}.
              {{question}}
              A. {{choice_a}}
              B. {{choice_b}}
              C. {{choice_c}}
              D. {{choice_d}}{% if choice_e is defined and choice_e %}
              E. {{choice_e}}{% endif %}"
\end{verbatim}
% UltraEval‑Audio 的 Prompt 管理机制通过配置驱动、模板复用与动态渲染实现了评测流程的统一与自动化。一方面，它彻底解耦了提示词设计与核心评测逻辑，降低了跨任务与跨模型的集成成本；另一方面，它确保了提示词定义的可重现性、可扩展性与可共享性，为后续音频大模型评测的标准化奠定了坚实基础。
The prompt management mechanism achieves unified and automated evaluation through configuration-driven design, template reuse, and dynamic rendering. On the one hand, it fully decouples prompt design from the core evaluation logic, reducing integration overhead across tasks and models. On the other hand, it ensures that prompt definitions are reproducible, extensible, and easily shareable, laying a solid foundation for standardized evaluation of audio foundation models.
\vspace{-1.0em}
\subsection{Model Deployment}
\vspace{-1.0em}
\label{sec:model_delpoyment} 
% 模型部署是连接数据与评测的关键环节。当前音频模型的运行形态大致分为两类：（1）远程 API 模型（Remote API），通过官方 SDK 或 HTTP API 接口完成远程推理；（2）本地开源模型（Locally Deployed Models），依赖本地硬件资源进行推理计算。两类模型在依赖环境、硬件配置及运行方式上的差异，使得统一评测框架的集成与管理面临显著工程挑战。
Model deployment is a critical link between data and evaluation. Audio models are typically deployed in one of two forms: (1) Remote API models, which perform inference via official SDKs or HTTP API endpoints, and (2) Locally deployed models, which rely on local hardware resources for inference. Differences in environment dependencies, hardware requirements, and execution modes between these two types pose significant engineering challenges for a unified evaluation framework.

%尤其对于本地开源模型而言，依赖冲突（Dependency Conflict） 是最常见的问题。不同模型可能需要彼此不兼容的深度学习框架、驱动程序或音频处理库。例如，音频质量评测模块（如 UTMOS、SIM、DNSMOS）以及 ASR‑WER 评测器通常依赖特定环境版本，与被评模型的依赖栈存在冲突风险。若在同一运行空间中混合加载不同模型，极易导致环境污染或推理崩溃，从而影响评测的可复现性。
This challenge is especially pronounced for locally deployed models, where dependency conflicts are a common issue. Different models may require incompatible versions of deep learning frameworks, drivers, or audio processing libraries. For instance, audio quality evaluation modules (such as UTMOS, SIM, DNSMOS) and ASR-WER evaluators often depend on specific environment versions that may conflict with the dependency stacks of the models under evaluation. Mixing multiple models in the same runtime environment may lead to contamination or inference failures, compromising the reproducibility of evaluation results.

%为解决这一问题，UltraEval‑Audio 在架构设计中引入了**隔离运行（Isolated Runtime）**机制，用于在系统层面确保模型的环境独立性与运行安全性。该机制的核心设计如下：
To address this challenge, UltraEval-Audio introduces an \textbf{Isolated Runtime} mechanism ensuring environment independence and safe execution at the system level. The core design features are as follows:
\begin{enumerate}[leftmargin=*]
    \item \textbf{Environment-level isolation}: Each evaluated model is allocated an independent virtual runtime environment with only the dependencies it requires. This setup is completely isolated from the main evaluation process, eliminating dependency conflicts at the source.
    \item \textbf{Subprocess-based model execution}: Models run as independent subprocesses in a daemonized manner, maintaining a persistent loaded state to support continuous inference and significantly reduce the overhead associated with frequent model loading.
    \item \textbf{Inter-process communication (IPC)}: The main evaluation process communicates with each model subprocess via system pipes, exchanging data and inference results securely and with low latency.
\end{enumerate}
%1. 环境级隔离：为每个被评模型创建独立的虚拟运行环境，仅安装其所需依赖，与主评测进程完全隔离，从源头上消除依赖冲突。
%2. 子进程执行模型：模型作为独立子进程（Isolated Subprocess）运行，并以常驻服务（daemon process）形式保持加载状态，支持持续推理，显著降低频繁加载模型的资源开销。
%3. 进程间通信（IPC）：主评测进程与模型子进程之间通过 Inter‑Process Communication 机制（ system pipes）交换数据与推理结果，确保通信安全与低延迟。

%在设计层面，此机制等效于一种“本地微服务化（Micro‑service style）”的推理架构。用户在使用时无需手动管理或安装额外依赖，系统会在内部自动创建、配置拟环境，极大地降低了大规模多模型评测的工程复杂度。此外，为统一不同类型模型的调用方式，框架在部署层提供了标准化的 .inference() 接口，用于接收 Prompt 管理模块生成的输入并返回推理结果。
At the design level, this mechanism effectively implements a microservice-style inference architecture. Users do not need to manually manage or install additional dependencies, and the system significantly reduces the engineering complexity of large-scale, multi-model evaluation. Furthermore, to standardize the interface across different model types, the framework provides a unified \textit{.inference()} method at the deployment layer, which accepts inputs from the prompt management module and returns the inference results.

%通过以上设计，UltraEval‑Audio 在系统层面实现了模型运行的彻底解耦与自动化管理。一方面，隔离机制消除了模型间依赖冲突与环境污染问题，保障了评测实验的可复现性与结果一致性；另一方面，统一的推理接口抽象使得评测流程与模型底层实现无关，降低了跨模型比较与扩展的技术门槛。整体而言，该模块为大规模音频模型评测提供了稳定、可扩展且透明的运行底座。
Through this design, UltraEval-Audio achieves full decoupling and automated management of model execution at the system level. On one hand, the isolated runtime eliminates dependency conflicts and environment contamination between models, ensuring reproducibility and consistency of evaluation results. On the other hand, the unified inference interface abstracts away underlying model implementations, lowering the technical barrier for cross-model comparison and framework extension. Overall, this module provides a stable, scalable, and transparent runtime foundation for large-scale audio model evaluation.

\vspace{-1.0em}
\subsection{Evaluation}
\vspace{-1.0em}
\label{sec:evaluation}
% 完成模型推理后，UltraEval‑Audio 进入评测执行阶段。该阶段旨在将模型的原始输出转化为可度量的结构化结果，从而实现多模型间的客观性能衡量。整个评测流程由两个核心环节构成：后处理（Post‑processing）与评估器（Evaluator）。前者负责对模型输出进行规范化解析与语义对齐，后者则根据任务特征与模态类型计算相应的评测指标。通过这两级结构，系统实现了从结果生成到指标计算的全自动化与标准化流程。
Following model inference, UltraEval-Audio proceeds to the evaluation phase, in which raw model outputs are converted into structured and quantifiable results to enable objective performance assessment across models. The evaluation workflow comprises two core components: post-processing and evaluator. The post-processing module standardizes and semantically aligns the model outputs, while the evaluator computes metrics according to task. This two level structure enables a fully automated and standardized pipeline from output generation to metric calculation.

\textbf{Post-processing.}
%在多模态音频评测中，模型输出常常包含与任务目标无关的前后缀信息、格式性噪声或非结构化文本，这会干扰指标计算的准确性。为保证输入评测模块的数据一致性与结构完整性，UltraEval‑Audio 设计了灵活的后处理机制，对模型输出进行多层级的解析与规范化。
In multimodal audio evaluation, model outputs often include extraneous prefixes or suffixes, formatting artifacts, or unstructured text unrelated to the task objectives, which can compromise accurate metric computation. To ensure consistent and well-structured inputs for evaluation, UltraEval-Audio employs a flexible post-processing mechanism that parses and standardizes outputs at multiple levels.

%该机制采用模块化与可组合（composable）设计理念，支持基于任务定义的自适应工作流（workflow）构建。系统内置多种标准后处理模块，包括：选项提取（Option Extraction）、是/否判断解析（Yes/No  Parser）、结构化字段解析（JSON Field Parser）等，它们可按需串联形成多步后处理管线适配复杂任务。
This mechanism adopts a modular and composable design, enabling adaptive workflows based on task specifications. The framework includes several built-in post-processing modules—such as \textit{Option Extraction}, \textit{Yes/No Parser}, and \textit{JSON Field Parser}—which can be sequentially combined to form multi-step pipelines for handling complex tasks.

\textbf{Evaluator.}
%Evaluator 模块接收经后处理规范化后的模型输出，负责计算最终的性能指标。UltraEval-Audio 将评测器抽象为两类，
The evaluator module processes model outputs that are standardized via post-processing and is responsible for computing the final performance metrics. UltraEval-Audio categorizes evaluators into two types:

\begin{itemize}[leftmargin=*]
%基于规则的评测器（Rule-based  Evaluators）：针对具有标准答案（Ground Truth）的任务，采用经典的算法指标进行度量。本框架内置了针对 ASR 任务的词错误率（WER）、针对语音翻译（AST）的 BLEU/METEOR、针对分类任务的准确率（Accuracy）等等
    \item \textbf{Rule-based Evaluators}: For tasks with reference answers, classical algorithmic metrics are applied. The framework includes WER for ASR tasks, BLEU for AST, accuracy for classification tasks as well as other similar measures.
    % 基于模型的评测器（Model-based Evaluators）：针对音频生成质量或开放式生成任务，集成了一系列预训练评估模型以模拟人类的一致性判断。这涵盖了衡量音色克隆相似度的说话人编码余弦相似度（SIM）、量化语音自然度与质量的 UTMOS 与 DNSMOS，以及基于 GPT的开放域问答评分器（LLM-as-a-Judge）。
    \item \textbf{Model-based Evaluators}: For audio generation quality or open-ended generative tasks, pretrained evaluation models are employed to approximate human judgment. This includes Speaker Similarity (SIM) for assessing timbre or voice cloning fidelity, UTMOS and DNSMOS for evaluating naturalness and perceptual quality of speech, and GPT-based open-domain QA scorers (LLM-as-a-Judge).
\end{itemize}

%所有 Evaluator 均通过统一接口注册与调度，其输出结果自动聚合至统一的结果分析。该设计既保证了不同任务指标间的兼容性，又支持新评测器的快速扩展，从而使框架具备高度的可插拔性与可维护性。
All evaluators are registered and orchestrated through a unified interface, with their outputs automatically aggregated into a consolidated results summary. This design ensures compatibility across metrics from different tasks while facilitating rapid integration of new evaluators, rendering the framework highly modular and maintainable.

%UltraEval‑Audio 的评测执行层通过标准化的后处理机制与多维度评估体系，实现了从模型输出到量化结果的端到端自动化评测流程。该设计有效提升了评测结果的一致性、透明性与复现性，为多模态音频基础模型的客观比较与持续演进提供了系统化的技术支撑。
By employing standardized post-processing and a multidimensional evaluation framework, UltraEval-Audio establishes an end-to-end automated pipeline from model outputs to quantifiable metrics. This design enhances consistency, transparency, and reproducibility, providing a systematic foundation for objective comparison and ongoing development of multimodal audio foundation models.
\vspace{-1.0em}
\section{Evaluation Results}
\vspace{-1.0em}
%通过UltraEval-Audio，研究者能够基于多类基准测试对音频大语言模型及音频编解码器开展全面评估。该框架提供了一套统一的解决方案，可用于系统化评测这些音频处理模型在多样化测试环境中的性能表现。
% With \textbf{UltraEval-Audio}, researchers can comprehensively evaluate audio foundation models and audio codecs across multiple benchmarks. The framework provides a unified solution for systematically assessing the performance of these audio-processing models in various testing environments.
\textbf{UltraEval-Audio} provides a unified solution for systematically assessing the performance of audio-processing models in various testing environments. In this section, we carefully select representative evaluation tasks and build three leaderboards: audio understanding, audio generation, and audio codec. Then we evaluate 13 leading audio foundation models and 9 audio codecs, introduced in Section~\ref{eval:models}.
%在本节，我们首先在4.1节介绍了参与榜单评测的模型，之后，分别在4.2介绍了音频理解榜单，在4.3介绍了音频生成榜单，在4.4介绍了音频编解码器榜单
We present the leaderboards of audio understanding, audio generation, and audio codec in Section~\ref{eval:audio understanding},~\ref{eval:audio generation} and~\ref{eval:audio codec} respectively. %Section~\ref{eval:audio understanding} then presents the \textbf{Audio Understanding Leaderboard}, Section~\ref{eval:audio generation} discusses the \textbf{Audio Generation Leaderboard}, and Section~\ref{eval:audio codec} describes the \textbf{Audio Codec Leaderboard}.

\begin{table*}
  \centering
  \small
  \renewcommand{\arraystretch}{1.2}
  \caption{Overview of audio foundation models participating in the evaluation}
  \begin{tabular}{lcccc}
     \toprule
    \textbf{Model} & \textbf{Institution} & \textbf{Type} & \textbf{Modality} (Input$\rightarrow$Output) & \textbf{Languages} \\
    \midrule
    \textbf{GPT-4o-Realtime} & OpenAI & Proprietary & Audio + Text $\rightarrow$ Audio + Text & Multilingual \\
    \textbf{Qwen3-Omni-30B-A3B-Instruct} & Alibaba & Open-Source & Audio + Text $\rightarrow$ Audio + Text & Multilingual \\
    \textbf{Qwen2.5-Omni} & Alibaba & Open-Source & Audio + Text $\rightarrow$ Audio + Text & English, Chinese \\
    \textbf{MiniCPM-o 2.6} & OpenBMB & Open-Source & Audio + Text $\rightarrow$ Audio + Text & English, Chinese \\
    \textbf{Kimi-Audio-7B-Instruct} & Moonshot & Open-Source & Audio + Text $\rightarrow$ Audio + Text & English, Chinese \\
    \textbf{Gemini-1.5-Flash} & Google & Proprietary & Audio + Text $\rightarrow$ Text & Multilingual \\
    \textbf{Gemini-1.5-Pro} & Google & Proprietary & Audio + Text $\rightarrow$ Text & Multilingual \\
    \textbf{Gemini-2.5-Flash} & Google & Proprietary & Audio + Text $\rightarrow$ Text & Multilingual \\
    \textbf{Gemini-2.5-Pro} & Google & Proprietary & Audio + Text $\rightarrow$ Text & Multilingual \\
    \textbf{Qwen2-Audio-7B} & Alibaba & Open-Source & Audio + Text $\rightarrow$ Text & Multilingual \\
    \textbf{Qwen2-Audio-7B-Instruct} & Alibaba & Open-Source & Audio + Text $\rightarrow$ Text & Multilingual \\
    \textbf{MiDaShengLM-7B} & Xiaomi & Open-Source & Audio + Text $\rightarrow$ Text & Multilingual \\
    \textbf{GLM-4-Voice} & Zhipu AI& Open-Source & Audio $\rightarrow$ Audio & English, Chinese \\
     \bottomrule
  \end{tabular}
  \label{tab: support audio foundation models}
\end{table*}

\vspace{-1.0em}
\subsection{Evaluated Models}
\vspace{-1.0em}
\label{eval:models}
%我们挑选了热门和新兴的音频大模型，编解码器来构建榜单,音频大模型在表1，音频编解码器包含了encodec-48k，chattts-DVAE， mimi， WavTokenizer-large-v2-75-tokens， WavTokenizer-large-40-tokens
% spark
% 被评估的语音基准模型在表2中进行展示，包括了领先的闭源模型如GPT-4o和Gemini 2.5，以及最新的开源模型如Qwen3-Omni和Kimi-Audio。
We evaluate emerging audio foundation models and audio codecs to build leaderboards.
The evaluated audio foundation models are shown in Table~\ref{tab: support audio foundation models}, including leading proprietary models such as GPT-4o-Realtime and Gemini-2.5-Pro, as well as the latest open-source models like Qwen3-Omni-30B-A3B-Instruct and Kimi-Audio-7B-Instruct. Audio codecs include {Encodec}~\citep{défossez2022high}, {ChatTTS-DVAE}\footnote{https://github.com/2noise/ChatTTS}, the {Mimi}~\citep{kyutai2024moshi} family,  WavTokenizer-large-speech-75token (denoted as {WavTokenizer-large-75})\footnote{https://huggingface.co/novateur/WavTokenizer-large-speech-75token} , WavTokenizer-large-unify-40token (denoted as {WavTokenizer-large-40})\footnote{https://huggingface.co/novateur/WavTokenizer-large-unify-40token}~\citep{ji2024wavtokenizer} and {Spark}~\citep{wang2025sparktts}.

% 重要地，每个模型推理尽可能使用自身官方prompt和参数。为了保证一个公平和一致的评测协议，我们不做prompt或者参数优化。
Several models design different prompts and parameters for specific tasks, for example, Kimi-Audio-7B-Instruct has specific prompts for ASR tasks. To replicate these results, we run the tasks using the prompts and parameters provided by the publisher. For tasks where the publisher has not provided these configurations, we use the official inference parameters and prompts for evaluation. Furthermore, we do not perform additional prompt or parameter optimizations, ensuring a fair and consistent evaluation protocol.

\vspace{-1.0em}
\subsection{Audio Understanding}
\vspace{-1.0em}
\label{eval:audio understanding}

\begin{table*}[h]
  \centering
  \small
  \renewcommand{\arraystretch}{1.2}
  \setlength{\tabcolsep}{2.5pt}
  \caption{The audio understanding leaderboard. Best results are in bold. The average score is computed as the mean of all available metric scores, where WER-based metrics use ($100-\text{WER}$) and other metrics (e.g., BLEU/Acc.) are unchanged.}
  \resizebox{\textwidth}{!}{
  \begin{tabular}{lcccccccccc}
    \toprule
    \multirow{5}{*}{\textbf{Model}} & \multicolumn{6}{c}{\textbf{ASR}} & \multicolumn{2}{c}{\textbf{AST}} & \textbf{EMO} & \multirow{5}{*}{\textbf{Avg. Score} ($\uparrow$)} \\
    \cmidrule(lr){2-7} \cmidrule(lr){8-9} \cmidrule(lr){10-10} 
    & \multicolumn{1}{c}{\shortstack{\textbf{LibriSpeech}\\dev-clean | dev-other\\ test-clean | test-other}}
    &\textbf{TED-LIUM} &\multicolumn{1}{c}{\shortstack{\textbf{CV-15}\\en | zh}} & \textbf{AISHELL-1} & \textbf{FLEURS} & \multicolumn{1}{c}{\shortstack{\textbf{Wenet}\\\textbf{-test-net}}} & \textbf{covost2-en2zh} & \textbf{covost2-zh2en} & \textbf{MELD} &  \\
   \cmidrule(lr){2-7} \cmidrule(lr){8-9} \cmidrule(lr){10-10} 
    & WER ($\downarrow$)    & WER ($\downarrow$)& WER | CER ($\downarrow$) & CER ($\downarrow$) &CER ($\downarrow$) & CER ($\downarrow$) & BLEU ($\uparrow$) & BLEU ($\uparrow$) & Acc. ($\uparrow$)   \\
    \midrule
    \textbf{GPT-4o-Realtime} & \shortstack{2.30 | 5.60 \\ 2.60 | 5.50} & 4.80 & 27.44 | 37.44 & 7.30 & 5.40 & 28.90 & 37.10 & 15.70 & 33.20 & 73.75 \\
    \textbf{Qwen3-Omni-30B-A3B-Instruct} & \shortstack{1.25 | 2.27 \\ 1.36 | 2.57} & 2.82 & \textbf{6.00} | \textbf{4.32} & 0.87 & 2.61 & \textbf{4.82} & 46.58 & \textbf{29.40} & 56.81 & \textbf{84.92} \\
    \textbf{Qwen2.5-Omni} & \shortstack{2.10 | 4.20 \\ 2.40 | 4.20}  & 4.70 & 8.70 | 5.20 & 1.10 & 4.60&  6.00 & 42.50 & 11.50 & 53.60 & 81.88 \\
    \textbf{MiniCPM-o 2.6} & \shortstack{1.60 | 3.40 \\ 1.70 | 4.40}  & 3.00 & 10.30 | 9.60 & 1.60 & 4.40 & 6.90 & \textbf{48.20} & 27.20 & 52.40 & 83.15 \\
    \textbf{Kimi-Audio-7B-Instruct} & \shortstack{\textbf{1.18 | 2.34}\\\textbf{ 1.28 | 2.44}}  & 2.96 & 7.09 | 5.72 & \textbf{0.60} & \textbf{2.53} &  5.55 & 36.61 & 18.30 & \textbf{59.23} & 83.27 \\
    \textbf{Gemini-1.5-Flash} & \shortstack{ 5.90 | 7.20 \\ 21.90 | 16.30}  & 6.90 & 208.00 | 84.37 & 9.00 & 85.90 &  279.90 & 33.40 & 8.20 & 45.20 & 27.80 \\
    \textbf{Gemini-1.5-Pro} & \shortstack{2.60 | 4.40 \\ 2.90 | 4.90} & 3.00 & 8.36 | 13.26 & 4.50 & 5.90 &  14.30 & 47.30 & 22.60 & 48.40 & 81.09 \\
    \textbf{Gemini-2.5-Flash} & \shortstack{3.73 | 6.71 \\ 3.28 | 12.03} & 3.53 & 46.76 | 36.15 & 6.40 & 6.45 & 126.07 & 3.67 & 10.61 & 51.53 & 62.67 \\
    \textbf{Gemini-2.5-Pro} & \shortstack{5.30 | 4.51 \\ 2.84 | 6.74} & \textbf{2.52} & 9.42 | 11.04 & 3.36 & 4.25 & 16.83 & 41.75 & 27.84 & 46.59 & 80.72 \\
    \textbf{Qwen2-Audio-7B} & \shortstack{1.57 | 3.50 \\1.60 | 3.88} & 3.43 & 8.67 | 7.03 & 1.52 & 5.89 & 8.09 & 45.30 & 24.84 & 42.87 & 82.14 \\
    \textbf{Qwen2-Audio-7B-Instruct} & \shortstack{ 2.90 | 5.50 \\ 3.10 | 5.70}  & 5.90 & 10.68 | 8.39 & 2.60 & 6.90 &  10.30 & 39.50 & 22.90 & 17.40 & 78.29 \\
    \textbf{MiDaShengLM-7B} & \shortstack{2.20 | 4.75 \\ 2.21 | 5.16} & 146.53 & 13.66 | 29.13 & 1.23 & 3.28 & 16.56 & 38.52 & 22.68 & 53.96 & 68.50 \\
    \bottomrule
  \end{tabular}
  }
  \footnotesize
  \raggedright
  \textbf{Notes: }
  WER/CER values can be greater than 100 when the total number of recognition errors exceeds the number of reference words/characters.
\label{tab: audio understanding leaderboard}
\end{table*}

%在音频理解方面，我们选择知名并且尽可能在各个音频模型的论文里都有被提到的数据集，最终我们在ASR任务上选择librispeech(en), tedlium(en), common voice 15(zh, en), fleurs(zh), WenetSpeech-test-ne(zh), AISHELL-1(zh)，翻译任务里选择covost2-zh2en，covost2-en2zh， 情感识别里选择MELD
For audio understanding, we select well-established and widely cited benchmarks that are commonly used in existing papers. Specifically, we use \textbf{LibriSpeech} (en), \textbf{TED-LIUM} (en), \textbf{Common Voice 15} (en/zh), \textbf{AISHELL-1} (zh), \textbf{FLEURS} (zh), \textbf{WenetSpeech-test-net} (zh) for ASR, \textbf{covost2-en2zh}, \textbf{covost2-zh2en} for AST, and \textbf{MELD} for emotion recognition (EMO). The final results are shown in Table~\ref{tab: audio understanding leaderboard}, from which we can make the following key observations:

%详细表现在附录1，任务级别的榜单在表2，显示了下面关键
%GPT不在保持领先地位，开源模型如MiniCPM-o 2.6 and Kimi-Audio-7B-Instruct表现优于它。从表3可见GPT在中文ASR上表现不佳，这是它落后的主要原因。
%Kimi-Audio-7B-Instruct在 ASR，EMO任务上都领先于MiniCPM-o 2.6，但是在AST落后
% Qwen2-Audio-7B-Instruct于自身的base模型Qwen2-Audio-7B对齐存在一定问题
(1) GPT-4o-Realtime faces strong competition in the field of audio understanding, with open-source models such as Qwen3-Omni-30B-A3B-Instruct, Kimi-Audio-7B-Instruct, MiniCPM-o 2.6 as well as proprietary models like Gemini-2.5-Pro, achieving superior performance in the evaluation. A key reason for this is GPT-4o-Realtime's relatively underwhelming performance on Chinese ASR benchmarks, especially when evaluated on datasets such as \textbf{Wenet-test-net} and \textbf{Common Voice 15} (CV-15).
    
(2) Qwen3-Omni-30B-A3B-Instruct demonstrates superior performance across all tasks, consistently delivering high-quality results in various domains. Kimi-Audio-7B-Instruct excels in ASR and EMO tasks but underperforms in AST, indicating improvement room in the latter area.

(3) For models from Gemini family, we observe that each generation's Flash model performs weaker than the Pro model. However, compared to Gemini 1.5, the Flash model of Gemini 2.5 performs better, while the Pro models show similar performance.
    
% (3) Qwen2-Audio-7B-Instruct exhibits a slight performance drop compared to its base model Qwen2-Audio-7B on ASR tasks, while performs better on AST and EMO tasks.

\vspace{-1.0em}
\subsection{Audio Generation}
\vspace{-1.0em}
\label{eval:audio generation} 

\begin{table*}[h]
  \centering
  \small
  \renewcommand{\arraystretch}{1.2}
  \setlength{\tabcolsep}{2.5pt}
  \caption{The audio generation leaderboard. Acoustic metrics (UTMOS | DNSMOS P.835 | DNSMOS P.808, scores range from 0 to 5) are evaluated on the generated audio responses from the speech tasks.  Best results are in bold.}
  \resizebox{\textwidth}{!}{
  \begin{tabular}{lccccccc}
  \toprule
 \multirow{2}{*}{\textbf{Models}}& \multicolumn{1}{c}{\shortstack{\textbf{Speech}\\ \textbf{WebQuestions}}} & \multicolumn{1}{c}{\shortstack{\textbf{Speech}\\ \textbf{TriviaQA}}} & \multicolumn{1}{c}{\shortstack{\textbf{Speech}\\ \textbf{AlpacaEval}}}& \shortstack{\textbf{Speech}\\ \textbf{CMMLU}} &\shortstack{\textbf{Speech}\\ \textbf{HSK}} & \textbf{Acoustics} & \multirow{2}{*}{\textbf{Avg. Score} ($\uparrow$)} \\ 
    \cmidrule(lr){2-6} \cmidrule(lr){7-7} 
    & {Acc. ($\uparrow$)} & {Acc. ($\uparrow$)}& {Acc. ($\uparrow$)}& {Acc. ($\uparrow$)}& {Acc. ($\uparrow$)} & Acoustics ($\uparrow$)\\
    \midrule
\textbf{GPT-4o-Realtime} & \textbf{51.60} & \textbf{69.70} & \textbf{74.00} & 70.05 & \textbf{98.69} &  4.29 | 3.44 | 4.26 & \textbf{74.00} \\
\textbf{Qwen3-Omni-30B-A3B-Instruct} & 51.50 & 55.27 & 67.97 & 47.83 & 40.27 &  \textbf{4.44} | 3.45 | 4.12 & 57.15 \\
\textbf{Qwen2.5-Omni} & 38.89 & 39.94 & 54.00 & \textbf{73.72} & 95.65 &  4.23 | \textbf{3.48} | \textbf{4.27} & 63.68 \\
\textbf{MiniCPM-o 2.6} & 40.00 & 40.20 & 51.00 & 51.37 & 80.68 &  4.12 | 3.39 | 4.02 & 56.69 \\
\textbf{Kimi-Audio-7B-Instruct} & 33.69 & 38.20 & 34.40 & 71.25 & 97.42 &  2.94 | 3.22 | 3.62 & 56.69 \\
\textbf{GLM-4-Voice} & 32.00 & 36.40 & 51.00 & 52.61 & 71.06 &  4.21 | 3.46 | 4.07 & 53.56 \\
\bottomrule
    \end{tabular}
    }
\raggedright Note: The average score is computed as the average of 6 scores: five speech-task scores and normalized acoustic scores. For acoustic scores (UTMOS | DNSMOS P.835 | DNSMOS P.808), each value (0--5) is multiplied by 20 to map to 0--100, then averaged to obtain the normalized acoustic score.
\label{tab: audio generation leaderboard}
\end{table*}

To build the audio generation leaderboard, we select \textbf{SpeechWebQuestions}, \textbf{SpeechTriviaQA}, and \textbf{SpeechAlpacaEval} for English capability evaluation. Our self-proposed benchmarks \textbf{SpeechCMMLU} and \textbf{SpeechHSK} are also used to assess Chinese audio generation capabilities. We first evaluate the models' performance on each dataset by answer accuracy (Acc.), then assess the acoustic quality of the generated audio.

To ensure better alignment with existing evaluation results, we adopt different evaluation settings for different datasets. Specifically: 
\begin{itemize}[leftmargin=*]
    \item For \textbf{SpeechWebQuestions} and \textbf{SpeechTriviaQA}, we follow the evaluation approach in~\citep{nachmani2023spoken} and~\citep{kyutai2024moshi}. We transcribe the model's spoken responses using Whisper-large-v3, and a response is considered correct if the ground-truth answer appears in the transcription.
    \item For \textbf{SpeechAlpacaEval}, we adopt the evaluation protocol of~\citep{zeng2024glm4voiceintelligenthumanlikeendtoend}, employing GPT-4o-mini to assess the quality of the transcribed responses. Responses are rated on a scale of 1 to 10, following the MT-Bench rubric~\citep{zheng2023judging}.
    \item For \textbf{SpeechCMMLU} and \textbf{SpeechHSK}, we employ Paraformer-zh to transcribe the generated audio. The transcriptions are then matched with the multiple-choice options to calculate accuracy.
\end{itemize}

We further evaluate acoustic values of the generated audio responses from the aforementioned benchmarks, employing UTMOS, DNSMOS P.835, and DNSMOS P.808 as metrics.
%注意，moshi,Mini-omni, llama-omni因为不支持中文而被排除
The performance of all models is summarized in Table~\ref{tab: audio generation leaderboard}, with key findings as follows:

(1) GPT-4o-Realtime performs best in audio generation, particularly excelling in English benchmarks. It consistently produces high-quality, accurate, and natural-sounding speech, making it one of the top performers;

(2) Qwen3-Omni-30B-A3B-Instruct and Qwen2.5-Omni outperform GPT-4o-Realtime in acoustic metrics, demonstrating superior sound quality. These models offer more nuanced audio generation, providing richer and more realistic speech outputs;

(3) Kimi-Audio-7B-Instruct underperforms in acoustic quality, with its generated speech lacking some naturalness and clarity. This suggests that improvements are needed to make the output sound more natural and human-like.

\vspace{-1.0em}
\subsection{Audio Codec}
\vspace{-1.0em}
\label{eval:audio codec}
%我们使用纯净语音语料库对音频编解码器进行评估，包括\textbf{LibriSpeech-dev-clean}（英文）、\textbf{LibriSpeech-test-clean}（英文）和\textbf{aishell-1}（中文），
%分别通过语音识别词错误率、音色相似度和语音质量三项任务进行测试。
We evaluate audio codecs using clean speech corpora, including \textbf{LibriSpeech-dev-clean} (en), \textbf{LibriSpeech-test-clean} (en), and \textbf{AISHELL-1} (zh). For each dataset, we use the ASR-WER/CER metric to measure the codec accuracy, SIM (short for similarity) to measure timbre fidelity and acoustic scores to measure the acoustic quality.
The results are shown in Table~\ref{tab: audio codec leaderboard}, from which we observe the following:

\begin{table*}[h]
  \centering
  \small
  \renewcommand{\arraystretch}{1.2}
  \setlength{\tabcolsep}{2pt}
  \caption{The audio codec leaderboard. The hyphen (-) indicates that UTMOS is not applicable to Chinese speech (AISHELL-1). Best results are in bold.}
  \resizebox{\textwidth}{!}{
  \begin{tabular}{lcccccccccc}
    \toprule
    \multirow{2}{*}{\textbf{Models}} & \multicolumn{3}{c}{\textbf{LibriSpeech-dev-clean}} & \multicolumn{3}{c}{\textbf{LibriSpeech-test-clean}} & \multicolumn{3}{c}{\textbf{AISHELL-1}} & \multirow{2}{*}{\textbf{Avg. Score} ($\uparrow$)} \\
    \cmidrule(lr){2-4} \cmidrule(lr){5-7} \cmidrule(lr){8-10} 
    & {ASR-WER} ($\downarrow$) & {SIM} ($\uparrow$) & {Acoustics} ($\uparrow$) & {ASR-WER} ($\downarrow$) & {SIM} ($\uparrow$) & {Acoustics} ($\uparrow$) & {ASR-CER} ($\downarrow$) & {SIM} ($\uparrow$) & {Acoustics} ($\uparrow$) &  \\
    \midrule
\textbf{Encodec-24k} & 4.56 & 59.40 & 1.58 | 3.12 | 2.36 & 4.32 & 59.40 & 1.57 | 3.12 | 2.36 & 13.95 & 47.48 & - | 2.93 | 2.03 & 65.24 \\
\textbf{Encodec-48k} & 3.85 & 65.53 & 1.52 | 2.88 | 2.42 & 3.80 & 66.00 & 1.48 | 2.87 | 2.40 & 6.85 & 68.78 & - | 2.79 | 2.21 & 69.59 \\
\textbf{ChatTTS-DVAE} & 7.49 & 34.83 & 1.30 | 2.66 | 2.11 & 6.75 & 36.21 & 1.29 | 2.64 | 2.12 & 32.36 & 32.36 & - | 2.24 | 1.57 & 52.86 \\
\textbf{Mimi (32bit)} & \textbf{2.04} & \textbf{92.18} & 3.83 | 2.87 | 2.44 & \textbf{1.96} & \textbf{92.68} & 3.84 | 2.92 | 2.49 & \textbf{2.82} & \textbf{84.80} & - | 2.43 | 1.89 & 80.96 \\
\textbf{Mimi (8bit)} & 2.76 & 72.15 & 3.52 | 2.78 | 2.37 & 2.83 & 73.13 & 3.53 | 2.83 | 2.43 & 6.82 & 60.63 & - | 2.42 | 2.04 & 72.72 \\
\textbf{Mimi-streaming (8bit)} & 6.76 & 54.02 & 1.65 | 2.78 | 2.37 & 6.19 & 54.32 & 1.63 | 2.83 | 2.43 & 19.62 & 40.67 & - | 2.42 | 2.04 & 61.37 \\
\textbf{WavTokenizer-large-75} & 4.31 & 69.97 & 4.01 | 3.64 | \textbf{3.26} & 4.05 & 68.15 & 4.00 | 3.63 | \textbf{3.27} & 8.97 & 64.27 & - | 3.11 | \textbf{2.85} & 76.67 \\
\textbf{WavTokenizer-large-40} & 8.13 & 60.26 & 3.78 | 3.70 | 3.13 & 7.73 & 56.63 & 3.77 | 3.70 | 3.16 & 25.52 & 49.21 & - | 3.13 | 2.50 & 69.18 \\
\textbf{Spark} & 2.39 & 79.94 & \textbf{4.18} | \textbf{3.85} | 3.24 & 2.53 & 79.53 & \textbf{4.18} | \textbf{3.83} | 3.24 & 3.66 & 74.76 & - | \textbf{3.63} | \textbf{2.85} & \textbf{82.29} \\
    \bottomrule
  \end{tabular}
  }
\vspace{2pt}
\begin{minipage}{\textwidth}
\footnotesize
\raggedright
Note: For acoustic scores we also use UTMOS, DNSMOS P.835, and DNSMOS P.808 metrics. To calculate the average score, for ASR-WER and ASR-CER, we calculate $100-\text{val}$. For acoustic scores, each available value (ranges from 0 to 5) is normalized by $20\times\mathrm{val}$ (mapping to 0--100), and the acoustic score is their average (the hyphen `-' is ignored). The final score is the average of 9 metric scores.
\end{minipage}

\label{tab: audio codec leaderboard}
\end{table*}

\begin{figure}[htb]
    \centering
    \includegraphics[width=0.5\linewidth]{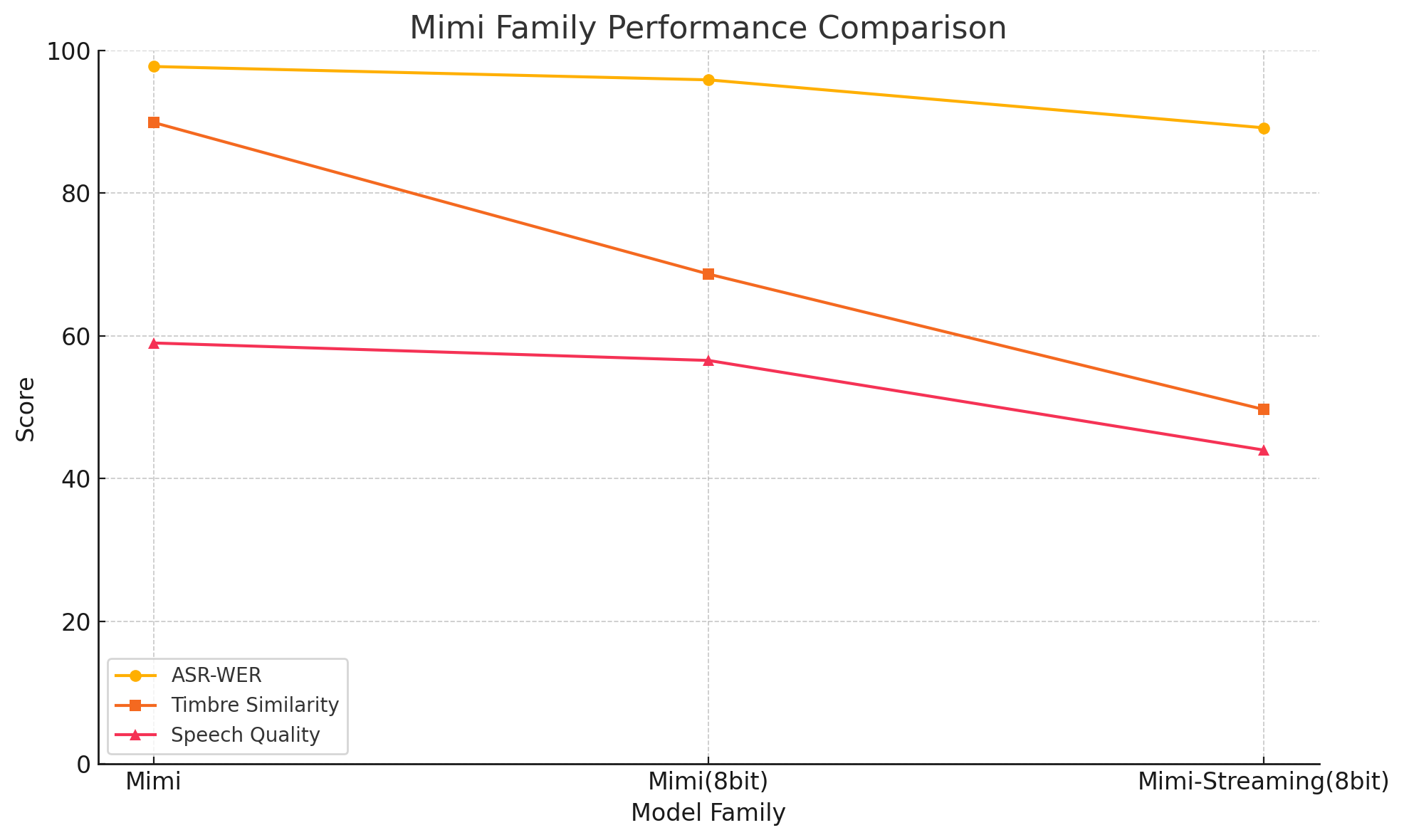}
    \caption{Mimi family performance comparison. Note that ASR-WER is normalized as ($100 - \text{WER}$), and speech quality scores (acoustic scores) are scaled by a factor of 20 for visualization. }
    \label{fig:mimi family}
\end{figure}
% 详细的性能结果见附录3， 各模型整体性能指标总结于表8，主要结论如下
% 各模型在语音识别词错误率方面的表现相对一致，而在音色相似度和语音质量方面则存在显著差异。后两项任务对编解码器性能的区分度更高。
% Mimi模型在语音识别词错误率和音色相似度上表现最佳，表明其token表征能有效捕捉原始音频中的语言信息和音色特征。然而，该模型的语音质量评分落后于Spark和WavTokenizer-large-v2-75-tokens，说明其解码器组件在合成自然度上尚有改进空间。
%  如图4所示，对比Mimi（默认32位）与Mimi（8位）版本，性能下降在音色相似度上最为显著，语音识别词错误率轻微下滑，语音质量适度下降。这表明音色信息在比特位上占据了更多的信息表示。流式变体模型则进一步降低了音色相似度和语音质量。
% 表9显示，ChatTTS-DVAE、WavTokenizer-large-40-tokens和Mimi-Streaming (8bit)在AISHELL-1数据集上表现不佳，表明这些模型需要改进对中文音频的处理能力。
(1) ASR-WER performance shows only limited disparity across models, whereas timbre fidelity and acoustic quality exhibit substantially larger variation. These latter dimensions provide more discriminative signals for assessing codec performance.
    
(2) The Mimi model performs best in codec accuracy and timbre fidelity, indicating that its token representation effectively captures both linguistic and timbral information from raw audio. However, its acoustic  scores lag behind Spark and WavTokenizer-large-75, suggesting that improvements could be made to its decoder.

(3) We further propose Figure~\ref{fig:mimi family} to compare Mimi codec variants. Comparing Mimi (default 32-bit) with Mimi (8bit), the performance drop is most pronounced in timbre fidelity, while ASR-WER slightly decreases, and acoustic quality drops modestly. This indicates that timbre information relies more heavily on higher bit depths. The streaming variant further degrades on both timbre fidelity and acoustic quality.

(4) ChatTTS-DVAE, WavTokenizer-large-40, and Mimi-Streaming (8bit) underperform on the AISHELL-1 dataset, indicating a need for improved handling of Chinese-language audio.

\vspace{-1.0em}
\section{Conclusion}
\vspace{-1.0em}

In this paper, we introduce UltraEval-Audio, the first unified evaluation framework for comprehensive assessment of audio foundation models. We first construct a complete audio evaluation taxonomy encompassing tasks and benchmarks. To enrich the evaluation, we develop a systematic methodology for assessing audio codecs and introduce two Chinese benchmarks: SpeechCMMLU and SpeechHSK. Building upon this, we implement a modular evaluation framework, which we then use to evaluate and analyze the performance of popular models across audio understanding, audio generation, and audio codec tasks.

% Limitations
\vspace{-1.0em}
\section{Limitations and Future Directions}
\vspace{-1.0em}
Our limitations are as follows. First, some current speech benchmarks rely on transcribed text as input for GPT-based evaluators rather than raw audio. This design introduces a dependency on ASR performance, which may propagate transcription errors into downstream judgments. Future work should therefore explore evaluation pipelines that operate directly on raw audio signals. In addition, the existing evaluation metrics are predominantly technical and do not adequately capture human perceptual factors, such as prosody, emotion, and whether the tone of the system's reply is appropriate for a given conversational context.

For future work, we plan to continuously update and refine the leaderboards, imporve inference capabilities (e.g. multi-GPU support), and incorporate evaluation methods that evaluate responses directly from raw audio. These enhancements will increase the comprehensiveness and reliability of audio foundation model evaluation, providing clearer guidance for the advancement of the field.

\newpage
\newpage
\bibliographystyle{citation}
%\bibstyle{unsrt}
\bibliography{citation}

@inproceedings{ye2025codec,
  title={Codec does matter: Exploring the semantic shortcoming of codec for audio language model},
  author={Ye, Zhen and Sun, Peiwen and Lei, Jiahe and Lin, Hongzhan and Tan, Xu and Dai, Zheqi and Kong, Qiuqiang and Chen, Jianyi and Pan, Jiahao and Liu, Qifeng and others},
  booktitle={Proceedings of the AAAI Conference on Artificial Intelligence},
  volume={39},
  number={24},
  pages={25697--25705},
  year={2025}
}

@article{anastassiou2024seed,
  title={Seed-tts: A family of high-quality versatile speech generation models},
  author={Anastassiou, Philip and Chen, Jiawei and Chen, Jitong and Chen, Yuanzhe and Chen, Zhuo and Chen, Ziyi and Cong, Jian and Deng, Lelai and Ding, Chuang and Gao, Lu and others},
  journal={arXiv preprint arXiv:2406.02430},
  year={2024}
}

@article{du2025cosyvoice,
  title={Cosyvoice 3: Towards in-the-wild speech generation via scaling-up and post-training},
  author={Du, Zhihao and Gao, Changfeng and Wang, Yuxuan and Yu, Fan and Zhao, Tianyu and Wang, Hao and Lv, Xiang and Wang, Hui and Ni, Chongjia and Shi, Xian and others},
  journal={arXiv preprint arXiv:2505.17589},
  year={2025}
}

@article{wang2025mgm,
  title={MGM-Omni: Scaling Omni LLMs to Personalized Long-Horizon Speech},
  author={Wang, Chengyao and Zhong, Zhisheng and Peng, Bohao and Yang, Senqiao and Liu, Yuqi and Gui, Haokun and Xia, Bin and Li, Jingyao and Yu, Bei and Jia, Jiaya},
  journal={arXiv preprint arXiv:2509.25131},
  year={2025}
}

@article{taal2011algorithm,
  title={An algorithm for intelligibility prediction of time--frequency weighted noisy speech},
  author={Taal, Cees H and Hendriks, Richard C and Heusdens, Richard and Jensen, Jesper},
  journal={IEEE Transactions on audio, speech, and language processing},
  volume={19},
  number={7},
  pages={2125--2136},
  year={2011},
  publisher={IEEE}
}

@article{series2014method,
  title={Method for the subjective assessment of intermediate quality level of audio systems},
  author={Series, B},
  journal={International Telecommunication Union Radiocommunication Assembly},
  volume={2},
  year={2014}
}

@article{hines2015visqol,
  title={ViSQOL: an objective speech quality model},
  author={Hines, Andrew and Skoglund, Jan and Kokaram, Anil C and Harte, Naomi},
  journal={EURASIP Journal on Audio, Speech, and Music Processing},
  volume={2015},
  number={1},
  pages={13},
  year={2015},
  publisher={Springer}
}

@inproceedings{chinen2020visqol,
  title={ViSQOL v3: An open source production ready objective speech and audio metric},
  author={Chinen, Michael and Lim, Felicia SC and Skoglund, Jan and Gureev, Nikita and O'Gorman, Feargus and Hines, Andrew},
  booktitle={2020 twelfth international conference on quality of multimedia experience (QoMEX)},
  pages={1--6},
  year={2020},
  organization={IEEE}
}

@article{lee2025ahelm,
  title={Ahelm: A holistic evaluation of audio-language models},
  author={Lee, Tony and Tu, Haoqin and Wong, Chi Heem and Wang, Zijun and Yang, Siwei and Mai, Yifan and Zhou, Yuyin and Xie, Cihang and Liang, Percy},
  journal={arXiv preprint arXiv:2508.21376},
  year={2025}
}

@article{surapaneni2025harness,
  title={AU-Harness: An Open-Source Toolkit for Holistic Evaluation of Audio LLMs},
  author={Surapaneni, Sidharth and Nguyen, Hoang and Mehta, Jash and Tiwari, Aman and Bamgbose, Oluwanifemi and Kalkunte, Akshay and Rajeswar, Sai and Madhusudhan, Sathwik Tejaswi},
  journal={arXiv preprint arXiv:2509.08031},
  year={2025}
}

@article{yao2024minicpm,
  title={MiniCPM-V: A GPT-4V Level MLLM on Your Phone},
  author={Yao, Yuan and Yu, Tianyu and Zhang, Ao and Wang, Chongyi and Cui, Junbo and Zhu, Hongji and Cai, Tianchi and Li, Haoyu and Zhao, Weilin and He, Zhihui and others},
  journal={arXiv preprint arXiv:2408.01800},
  year={2024}
}

@misc{openai2024gpt4ocard,
      title={GPT-4o System Card}, 
      author={OpenAI and : and Aaron Hurst and Adam Lerer and Adam P. Goucher and Adam Perelman and ... others},
      year={2024},
      eprint={2410.21276},
      archivePrefix={arXiv},
      primaryClass={cs.CL},
      url={https://arxiv.org/abs/2410.21276}, 
}

@article{xu2025qwen2,
  title={Qwen2. 5-omni technical report},
  author={Xu, Jin and Guo, Zhifang and He, Jinzheng and Hu, Hangrui and He, Ting and Bai, Shuai and Chen, Keqin and Wang, Jialin and Fan, Yang and Dang, Kai and others},
  journal={arXiv preprint arXiv:2503.20215},
  year={2025}
}

@article{ding2025kimi,
  title={Kimi-Audio Technical Report},
  author={Ding, Ding and Ju, Zeqian and Leng, Yichong and Liu, Songxiang and Liu, Tong and Shang, Zeyu and Shen, Kai and Song, Wei and Tan, Xu and Tang, Heyi and others},
  journal={arXiv preprint arXiv:2504.18425},
  year={2025}
}

@article{yang2024air,
  title={Air-bench: Benchmarking large audio-language models via generative comprehension},
  author={Yang, Qian and Xu, Jin and Liu, Wenrui and Chu, Yunfei and Jiang, Ziyue and Zhou, Xiaohuan and Leng, Yichong and Lv, Yuanjun and Zhao, Zhou and Zhou, Chang and others},
  journal={arXiv preprint arXiv:2402.07729},
  year={2024}
}

@article{zheng2023judging,
  title={Judging llm-as-a-judge with mt-bench and chatbot arena},
  author={Zheng, Lianmin and Chiang, Wei-Lin and Sheng, Ying and Zhuang, Siyuan and Wu, Zhanghao and Zhuang, Yonghao and Lin, Zi and Li, Zhuohan and Li, Dacheng and Xing, Eric and others},
  journal={Advances in Neural Information Processing Systems},
  volume={36},
  pages={46595--46623},
  year={2023}
}

@article{nachmani2023spoken,
  title={Spoken question answering and speech continuation using spectrogram-powered llm},
  author={Nachmani, Eliya and Levkovitch, Alon and Hirsch, Roy and Salazar, Julian and Asawaroengchai, Chulayuth and Mariooryad, Soroosh and Rivlin, Ehud and Skerry-Ryan, RJ and Ramanovich, Michelle Tadmor},
  journal={arXiv preprint arXiv:2305.15255},
  year={2023}
}

@inproceedings{reddy2022dnsmos,
  title={DNSMOS P. 835: A non-intrusive perceptual objective speech quality metric to evaluate noise suppressors},
  author={Reddy, Chandan KA and Gopal, Vishak and Cutler, Ross},
  booktitle={ICASSP 2022-2022 IEEE international conference on acoustics, speech and signal processing (ICASSP)},
  pages={886--890},
  year={2022},
  organization={IEEE}
}

@inproceedings{duan2024vlmevalkit,
  title={Vlmevalkit: An open-source toolkit for evaluating large multi-modality models},
  author={Duan, Haodong and Yang, Junming and Qiao, Yuxuan and Fang, Xinyu and Chen, Lin and Liu, Yuan and Dong, Xiaoyi and Zang, Yuhang and Zhang, Pan and Wang, Jiaqi and others},
  booktitle={Proceedings of the 32nd ACM international conference on multimedia},
  pages={11198--11201},
  year={2024}
}

@misc{2023opencompass,
    title={OpenCompass: A Universal Evaluation Platform for Foundation Models},
    author={OpenCompass Contributors},
    howpublished = {\url{https://github.com/open-compass/opencompass}},
    year={2023}
}

@techreport{kyutai2024moshi,
      title={Moshi: a speech-text foundation model for real-time dialogue},
      author={Alexandre D\'efossez and Laurent Mazar\'e and Manu Orsini and
      Am\'elie Royer and Patrick P\'erez and Herv\'e J\'egou and Edouard Grave and Neil Zeghidour},
      year={2024},
      eprint={2410.00037},
      archivePrefix={arXiv},
      primaryClass={eess.AS},
      url={https://arxiv.org/abs/2410.00037},
}

@article{joshi2017triviaqa,
  title={Triviaqa: A large scale distantly supervised challenge dataset for reading comprehension},
  author={Joshi, Mandar and Choi, Eunsol and Weld, Daniel S and Zettlemoyer, Luke},
  journal={arXiv preprint arXiv:1705.03551},
  year={2017}
}

@article{chen2024voicebench,
  title={Voicebench: Benchmarking llm-based voice assistants},
  author={Chen, Yiming and Yue, Xianghu and Zhang, Chen and Gao, Xiaoxue and Tan, Robby T and Li, Haizhou},
  journal={arXiv preprint arXiv:2410.17196},
  year={2024}
}

@misc{saeki2022utmosutokyosarulabvoicemoschallenge,
      title={UTMOS: UTokyo-SaruLab System for VoiceMOS Challenge 2022}, 
      author={Takaaki Saeki and Detai Xin and Wataru Nakata and Tomoki Koriyama and Shinnosuke Takamichi and Hiroshi Saruwatari},
      year={2022},
      eprint={2204.02152},
      archivePrefix={arXiv},
      primaryClass={cs.SD},
      url={https://arxiv.org/abs/2204.02152}, 
}

@misc{fang2025llamaomniseamlessspeechinteraction,
      title={LLaMA-Omni: Seamless Speech Interaction with Large Language Models}, 
      author={Qingkai Fang and Shoutao Guo and Yan Zhou and Zhengrui Ma and Shaolei Zhang and Yang Feng},
      year={2025},
      eprint={2409.06666},
      archivePrefix={arXiv},
      primaryClass={cs.CL},
      url={https://arxiv.org/abs/2409.06666}, 
}

@misc{zeng2024glm4voiceintelligenthumanlikeendtoend,
      title={GLM-4-Voice: Towards Intelligent and Human-Like End-to-End Spoken Chatbot}, 
      author={Aohan Zeng and Zhengxiao Du and Mingdao Liu and Kedong Wang and Shengmin Jiang and Lei Zhao and Yuxiao Dong and Jie Tang},
      year={2024},
      eprint={2412.02612},
      archivePrefix={arXiv},
      primaryClass={cs.CL},
      url={https://arxiv.org/abs/2412.02612}, 
}

@misc{geminiteam2024gemini15unlockingmultimodal,
      title={Gemini 1.5: Unlocking multimodal understanding across millions of tokens of context}, 
      author={Gemini Team and Petko Georgiev and Ving Ian Lei and ... others},
      year={2024},
      eprint={2403.05530},
      archivePrefix={arXiv},
      primaryClass={cs.CL},
      url={https://arxiv.org/abs/2403.05530}, 
}

@misc{chu2023qwenaudioadvancinguniversalaudio,
      title={Qwen-Audio: Advancing Universal Audio Understanding via Unified Large-Scale Audio-Language Models}, 
      author={Yunfei Chu and Jin Xu and Xiaohuan Zhou and Qian Yang and Shiliang Zhang and Zhijie Yan and Chang Zhou and Jingren Zhou},
      year={2023},
      eprint={2311.07919},
      archivePrefix={arXiv},
      primaryClass={eess.AS},
      url={https://arxiv.org/abs/2311.07919}, 
}

@misc{chu2024qwen2audiotechnicalreport,
      title={Qwen2-Audio Technical Report}, 
      author={Yunfei Chu and Jin Xu and Qian Yang and Haojie Wei and Xipin Wei and Zhifang Guo and Yichong Leng and Yuanjun Lv and Jinzheng He and Junyang Lin and Chang Zhou and Jingren Zhou},
      year={2024},
      eprint={2407.10759},
      archivePrefix={arXiv},
      primaryClass={eess.AS},
      url={https://arxiv.org/abs/2407.10759}, 
}

@misc{huang2025stepaudiounifiedunderstandinggeneration,
      title={Step-Audio: Unified Understanding and Generation in Intelligent Speech Interaction}, 
      author={Ailin Huang and Boyong Wu and Bruce Wang and Chao Yan and Chen Hu and ... others},
      year={2025},
      eprint={2502.11946},
      archivePrefix={arXiv},
      primaryClass={cs.CL},
      url={https://arxiv.org/abs/2502.11946}, 
}

@article{chen2015microsoft,
  title={Microsoft coco captions: Data collection and evaluation server},
  author={Chen, Xinlei and Fang, Hao and Lin, Tsung-Yi and Vedantam, Ramakrishna and Gupta, Saurabh and Doll{\'a}r, Piotr and Zitnick, C Lawrence},
  journal={arXiv preprint arXiv:1504.00325},
  year={2015}
}

@misc{li2023cmmlu,
      title={CMMLU: Measuring massive multitask language understanding in Chinese}, 
      author={Haonan Li and Yixuan Zhang and Fajri Koto and Yifei Yang and Hai Zhao and Yeyun Gong and Nan Duan and Timothy Baldwin},
      year={2023},
      eprint={2306.09212},
      archivePrefix={arXiv},
      primaryClass={cs.CL}
}

@misc{défossez2022high,
      title={High Fidelity Neural Audio Compression}, 
      author={Alexandre Défossez and Jade Copet and Gabriel Synnaeve and Yossi Adi},
      year={2022},
      eprint={2210.13438},
      archivePrefix={arXiv},
      primaryClass={eess.AS}
}

@article{ji2024wavtokenizer,
  title={Wavtokenizer: an efficient acoustic discrete codec tokenizer for audio language modeling},
  author={Ji, Shengpeng and Jiang, Ziyue and Wang, Wen and Chen, Yifu and Fang, Minghui and Zuo, Jialong and Yang, Qian and Cheng, Xize and Wang, Zehan and Li, Ruiqi and others},
  journal={arXiv preprint arXiv:2408.16532},
  year={2024}
}

@misc{wang2025sparktts,
      title={Spark-TTS: An Efficient LLM-Based Text-to-Speech Model with Single-Stream Decoupled Speech Tokens}, 
      author={Xinsheng Wang and Mingqi Jiang and Ziyang Ma and Ziyu Zhang and Songxiang Liu and Linqin Li and Zheng Liang and Qixi Zheng and Rui Wang and Xiaoqin Feng and Weizhen Bian and Zhen Ye and Sitong Cheng and Ruibin Yuan and Zhixian Zhao and Xinfa Zhu and Jiahao Pan and Liumeng Xue and Pengcheng Zhu and Yunlin Chen and Zhifei Li and Xie Chen and Lei Xie and Yike Guo and Wei Xue},
      year={2025},
      eprint={2503.01710},
      archivePrefix={arXiv},
      primaryClass={cs.SD},
      url={https://arxiv.org/abs/2503.01710}, 
}

@article{gulati2020conformer,
  title={Conformer: Convolution-augmented transformer for speech recognition},
  author={Gulati, Anmol and Qin, James and Chiu, Chung-Cheng and Parmar, Niki and Zhang, Yu and Yu, Jiahui and Han, Wei and Wang, Shibo and Zhang, Zhengdong and Wu, Yonghui and others},
  journal={arXiv preprint arXiv:2005.08100},
  year={2020}
}

@misc{alpaca_eval,
  author = {Xuechen Li and Tianyi Zhang and Yann Dubois and Rohan Taori and Ishaan Gulrajani and Carlos Guestrin and Percy Liang and Tatsunori B. Hashimoto },
  title = {AlpacaEval: An Automatic Evaluator of Instruction-following Models},
  year = {2023},
  month = {5},
  publisher = {GitHub},
  journal = {GitHub repository},
  howpublished = {\url{https://github.com/tatsu-lab/alpaca_eval}}
}

@misc{zhou2023instructionfollowingevaluationlargelanguage,
      title={Instruction-Following Evaluation for Large Language Models}, 
      author={Jeffrey Zhou and Tianjian Lu and Swaroop Mishra and Siddhartha Brahma and Sujoy Basu and Yi Luan and Denny Zhou and Le Hou},
      year={2023},
      eprint={2311.07911},
      archivePrefix={arXiv},
      primaryClass={cs.CL},
      url={https://arxiv.org/abs/2311.07911}, 
}

@misc{pascal-chsc-2011, 
author = "Bentley, P. and Nordehn, G. and Coimbra, M. and Mannor, S.", 
title = "The {PASCAL} {C}lassifying {H}eart {S}ounds {C}hallenge 2011 {(CHSC2011)} {R}esults", howpublished = "http://www.peterjbentley.com/heartchallenge/index.html"}

@INPROCEEDINGS{gong_vocalsound,
  author={Gong, Yuan and Yu, Jin and Glass, James},
  booktitle={ICASSP 2022 - 2022 IEEE International Conference on Acoustics, Speech and Signal Processing (ICASSP)}, 
  title={Vocalsound: A Dataset for Improving Human Vocal Sounds Recognition}, 
  year={2022},
  pages={151-155},
  doi={10.1109/ICASSP43922.2022.9746828}}

@article{Pratap2020MLSAL,
  title={MLS: A Large-Scale Multilingual Dataset for Speech Research},
  author={Vineel Pratap and Qiantong Xu and Anuroop Sriram and Gabriel Synnaeve and Ronan Collobert},
  journal={ArXiv},
  year={2020},
  volume={abs/2012.03411}
}

@article{fleurs2022arxiv,
  title = {FLEURS: Few-shot Learning Evaluation of Universal Representations of Speech},
  author = {Conneau, Alexis and Ma, Min and Khanuja, Simran and Zhang, Yu and Axelrod, Vera and Dalmia, Siddharth and Riesa, Jason and Rivera, Clara and Bapna, Ankur},
  journal={arXiv preprint arXiv:2205.12446},
  url = {https://arxiv.org/abs/2205.12446},
  year = {2022},
}

@misc{wang2020covost,
    title={CoVoST 2: A Massively Multilingual Speech-to-Text Translation Corpus},
    author={Changhan Wang and Anne Wu and Juan Pino},
    year={2020},
    eprint={2007.10310},
    archivePrefix={arXiv},
    primaryClass={cs.CL}
}

@inproceedings{drossos2020clotho,
  title={Clotho: An audio captioning dataset},
  author={Drossos, Konstantinos and Lipping, Samuel and Virtanen, Tuomas},
  booktitle={ICASSP 2020-2020 IEEE International Conference on Acoustics, Speech and Signal Processing (ICASSP)},
  pages={736--740},
  year={2020},
  organization={IEEE}
}

@inproceedings{turpault2019sound,
  title={Sound event detection in domestic environments with weakly labeled data and soundscape synthesis},
  author={Turpault, Nicolas and Serizel, Romain and Shah, Ankit Parag and Salamon, Justin},
  booktitle={Workshop on Detection and Classification of Acoustic Scenes and Events},
  year={2019}
}

@inproceedings{rocha2018alpha,
  title={A respiratory sound database for the development of automated classification},
  author={Rocha, BM and Filos, Dimitris and Mendes, Lea and Vogiatzis, Ioannis and Perantoni, Eleni and Kaimakamis, Evangelos and Natsiavas, P and Oliveira, Ana and J{\'a}come, C and Marques, A and others},
  booktitle={Precision Medicine Powered by pHealth and Connected Health: ICBHI 2017, Thessaloniki, Greece, 18-21 November 2017},
  pages={33--37},
  year={2018},
  organization={Springer}
}

@article{sakshi2024mmau,
  title={Mmau: A massive multi-task audio understanding and reasoning benchmark},
  author={Sakshi, S and Tyagi, Utkarsh and Kumar, Sonal and Seth, Ashish and Selvakumar, Ramaneswaran and Nieto, Oriol and Duraiswami, Ramani and Ghosh, Sreyan and Manocha, Dinesh},
  journal={arXiv preprint arXiv:2410.19168},
  year={2024}
}

@inproceedings{rousseau2012ted,
  title={TED-LIUM: an Automatic Speech Recognition dedicated corpus.},
  author={Rousseau, Anthony and Del{\'e}glise, Paul and Esteve, Yannick},
  booktitle={LREC},
  pages={125--129},
  year={2012}
}

@article{wang2021voxpopuli,
  title={VoxPopuli: A large-scale multilingual speech corpus for representation learning, semi-supervised learning and interpretation},
  author={Wang, Changhan and Riviere, Morgane and Lee, Ann and Wu, Anne and Talnikar, Chaitanya and Haziza, Daniel and Williamson, Mary and Pino, Juan and Dupoux, Emmanuel},
  journal={arXiv preprint arXiv:2101.00390},
  year={2021}
}

@inproceedings{tang2021kespeech,
  title={Kespeech: An open source speech dataset of mandarin and its eight subdialects},
  author={Tang, Zhiyuan and Wang, Dong and Xu, Yanguang and Sun, Jianwei and Lei, Xiaoning and Zhao, Shuaijiang and Wen, Cheng and Tan, Xingjun and Xie, Chuandong and Zhou, Shuran and others},
  booktitle={Thirty-fifth Conference on Neural Information Processing Systems Datasets and Benchmarks Track (Round 2)},
  year={2021}
}

@inproceedings{panayotov2015librispeech,
  title={Librispeech: an asr corpus based on public domain audio books},
  author={Panayotov, Vassil and Chen, Guoguo and Povey, Daniel and Khudanpur, Sanjeev},
  booktitle={2015 IEEE international conference on acoustics, speech and signal processing (ICASSP)},
  pages={5206--5210},
  year={2015},
  organization={IEEE}
}

@article{galvez2021people,
  title={The people's speech: A large-scale diverse english speech recognition dataset for commercial usage},
  author={Galvez, Daniel and Diamos, Greg and Ciro, Juan and Cer{\'o}n, Juan Felipe and Achorn, Keith and Gopi, Anjali and Kanter, David and Lam, Maximilian and Mazumder, Mark and Reddi, Vijay Janapa},
  journal={arXiv preprint arXiv:2111.09344},
  year={2021}
}

@inproceedings{zhang2022wenetspeech,
  title={Wenetspeech: A 10000+ hours multi-domain mandarin corpus for speech recognition},
  author={Zhang, Binbin and Lv, Hang and Guo, Pengcheng and Shao, Qijie and Yang, Chao and Xie, Lei and Xu, Xin and Bu, Hui and Chen, Xiaoyu and Zeng, Chenchen and others},
  booktitle={ICASSP 2022-2022 IEEE International Conference on Acoustics, Speech and Signal Processing (ICASSP)},
  pages={6182--6186},
  year={2022},
  organization={IEEE}
}

@article{chen2021gigaspeech,
  title={Gigaspeech: An evolving, multi-domain asr corpus with 10,000 hours of transcribed audio},
  author={Chen, Guoguo and Chai, Shuzhou and Wang, Guanbo and Du, Jiayu and Zhang, Wei-Qiang and Weng, Chao and Su, Dan and Povey, Daniel and Trmal, Jan and Zhang, Junbo and others},
  journal={arXiv preprint arXiv:2106.06909},
  year={2021}
}

@inproceedings{bu2017aishell,
  title={Aishell-1: An open-source mandarin speech corpus and a speech recognition baseline},
  author={Bu, Hui and Du, Jiayu and Na, Xingyu and Wu, Bengu and Zheng, Hao},
  booktitle={2017 20th conference of the oriental chapter of the international coordinating committee on speech databases and speech I/O systems and assessment (O-COCOSDA)},
  pages={1--5},
  year={2017},
  organization={IEEE}
}

@article{ardila2019common,
  title={Common voice: A massively-multilingual speech corpus},
  author={Ardila, Rosana and Branson, Megan and Davis, Kelly and Henretty, Michael and Kohler, Michael and Meyer, Josh and Morais, Reuben and Saunders, Lindsay and Tyers, Francis M and Weber, Gregor},
  journal={arXiv preprint arXiv:1912.06670},
  year={2019}
}

@article{dupuis2010toronto,
  title={Toronto emotional speech set (tess)-younger talker\_happy},
  author={Dupuis, Kate and Pichora-Fuller, M Kathleen},
  year={2010},
  publisher={Toronto: University of Toronto, Psychology Department, 2010.}
}

@article{poria2018meld,
  title={Meld: A multimodal multi-party dataset for emotion recognition in conversations},
  author={Poria, Soujanya and Hazarika, Devamanyu and Majumder, Navonil and Naik, Gautam and Cambria, Erik and Mihalcea, Rada},
  journal={arXiv preprint arXiv:1810.02508},
  year={2018}
}

@article{nagrani2017voxceleb,
  title={Voxceleb: a large-scale speaker identification dataset},
  author={Nagrani, Arsha and Chung, Joon Son and Zisserman, Andrew},
  journal={arXiv preprint arXiv:1706.08612},
  year={2017}
}

@inproceedings{kim2019audiocaps,
  title={Audiocaps: Generating captions for audios in the wild},
  author={Kim, Chris Dongjoo and Kim, Byeongchang and Lee, Hyunmin and Kim, Gunhee},
  booktitle={Proceedings of the 2019 Conference of the North American Chapter of the Association for Computational Linguistics: Human Language Technologies, Volume 1 (Long and Short Papers)},
  pages={119--132},
  year={2019}
}

@article{mei2024wavcaps,
  title={Wavcaps: A chatgpt-assisted weakly-labelled audio captioning dataset for audio-language multimodal research},
  author={Mei, Xinhao and Meng, Chutong and Liu, Haohe and Kong, Qiuqiang and Ko, Tom and Zhao, Chengqi and Plumbley, Mark D and Zou, Yuexian and Wang, Wenwu},
  journal={IEEE/ACM Transactions on Audio, Speech, and Language Processing},
  year={2024},
  publisher={IEEE}
}

@inproceedings{engel2017neural,
  title={Neural audio synthesis of musical notes with wavenet autoencoders},
  author={Engel, Jesse and Resnick, Cinjon and Roberts, Adam and Dieleman, Sander and Norouzi, Mohammad and Eck, Douglas and Simonyan, Karen},
  booktitle={International conference on machine learning},
  pages={1068--1077},
  year={2017},
  organization={PMLR}
}

@article{sturm2013gtzan,
  title={The GTZAN dataset: Its contents, its faults, their effects on evaluation, and its future use},
  author={Sturm, Bob L},
  journal={arXiv preprint arXiv:1306.1461},
  year={2013}
}

@misc{deepcontractor2023chord,
  author       = {deepcontractor},
  title        = {Musical Instrument Chord Classification},
  year         = {2023},
  howpublished = {\url{https://www.kaggle.com/datasets/deepcontractor/musical-instrument-chord-classification}},
  note         = {Accessed: 2025-05-13}
}

@misc{mmoreaux2023catsdogs,
  author       = {Mathieu Moreaux},
  title        = {Audio Cats and Dogs},
  year         = {2023},
  howpublished = {\url{https://www.kaggle.com/datasets/mmoreaux/audio-cats-and-dogs}},
  note         = {Accessed: 2025-05-13}
}

@misc{sripaadsrinivasan2023audiomnist,
  author       = {Sripaa D. Srinivasan},
  title        = {Audio MNIST},
  year         = {2023},
  howpublished = {\url{https://www.kaggle.com/datasets/sripaadsrinivasan/audio-mnist}},
  note         = {Accessed: 2025-05-13}
}

@article{dong2020interactive,
  title={An interactive web-based dashboard to track COVID-19 in real time},
  author={Dong, Ensheng and Du, Hongru and Gardner, Lauren},
  journal={The Lancet infectious diseases},
  volume={20},
  number={5},
  pages={533--534},
  year={2020},
  publisher={Elsevier}
}

@inproceedings{reddy2021dnsmos,
  title={DNSMOS: A non-intrusive perceptual objective speech quality metric to evaluate noise suppressors},
  author={Reddy, Chandan KA and Gopal, Vishak and Cutler, Ross},
  booktitle={ICASSP 2021-2021 IEEE International Conference on Acoustics, Speech and Signal Processing (ICASSP)},
  pages={6493--6497},
  year={2021},
  organization={IEEE}
}

@misc{he2024ultraeval,
      title={UltraEval: A Lightweight Platform for Flexible and Comprehensive Evaluation for LLMs}, 
      author={Chaoqun He and Renjie Luo and Shengding Hu and Yuanqian Zhao and Jie Zhou and Hanghao Wu and Jiajie Zhang and Xu Han and Zhiyuan Liu and Maosong Sun},
      year={2024},
      eprint={2404.07584},
      archivePrefix={arXiv},
      primaryClass={cs.CL}
}

@article{zeghidour2021soundstream,
  title={Soundstream: An end-to-end neural audio codec},
  author={Zeghidour, Neil and Luebs, Alejandro and Omran, Ahmed and Skoglund, Jan and Tagliasacchi, Marco},
  journal={IEEE/ACM Transactions on Audio, Speech, and Language Processing},
  volume={30},
  pages={495--507},
  year={2021},
  publisher={IEEE}
}

@article{kumar2023high,
  title={High-fidelity audio compression with improved rvqgan},
  author={Kumar, Rithesh and Seetharaman, Prem and Luebs, Alejandro and Kumar, Ishaan and Kumar, Kundan},
  journal={Advances in Neural Information Processing Systems},
  volume={36},
  pages={27980--27993},
  year={2023}
}

@article{yang2023hifi,
  title={Hifi-codec: Group-residual vector quantization for high fidelity audio codec},
  author={Yang, Dongchao and Liu, Songxiang and Huang, Rongjie and Tian, Jinchuan and Weng, Chao and Zou, Yuexian},
  journal={arXiv preprint arXiv:2305.02765},
  year={2023}
}

@article{xin2024bigcodec,
  title={Bigcodec: Pushing the limits of low-bitrate neural speech codec},
  author={Xin, Detai and Tan, Xu and Takamichi, Shinnosuke and Saruwatari, Hiroshi},
  journal={arXiv preprint arXiv:2409.05377},
  year={2024}
}

\clearpage

\end{document}